\documentclass[structabstract]{aa}  
\usepackage{graphicx}
\usepackage{txfonts}
\usepackage{longtable}
\begin{document}
   \title{Complex molecule formation in grain mantles}


   \author{P. Hall\inst{1, 2}
          \and
          T. J. Millar\inst{3}
          }

   \offprints{P. Hall \\ \email{phall@isc.astro.cornell.edu}}

   \institute{Department of Physics, UMIST, P.O. Box 88,
             Manchester, M60 1QD, UK.
         \and
              Astronomy Department, Cornell University,
              Ithaca, New York, 14853-6801, USA.
         \and
             Astrophysics Research Centre, School of Mathematics and Physics, Queen's University,
             Belfast, BT7 1NN, Northern Ireland.
            }

   \date{Preprint online version : 31 March 2010}

 
  \abstract
   {Complex molecules such as ethanol and dimethyl ether have 
    been observed in a number of hot molecular cores and hot corinos.
    Attempts to model the molecular formation process using gas
    phase only models have so far been unsuccessful.}
   {To demonstrate that grain surface processing is a 
    viable mechanism for complex molecule formation in these environments.}
   {A variable environment parameter computer model has been constructed
    which includes both gas and surface chemistry.
    This is used to investigate a variety of cloud collapse scenarios.}
   {Comparison between model results and observation shows that by combining 
    grain surface processing with gas phase chemistry complex molecules can be produced 
    in observed abundances in a number of core and corino scenarios.
    Differences in abundances are due to the initial atomic and molecular
    composition of the core/corino and varying collapse timescales.}
   {Grain surface processing, combined with variation of physical conditions,
    can be regarded as a viable method for the formation of complex molecules  
    in the environment found in the vicinity of a hot core/corino and produce 
    abundances comparable to those observed.}


   \maketitle


\section{Introduction}

A number of complex molecules have been discovered in the interstellar 
medium. First was methanol, CH$_{3}$OH, (Ball \cite{Ball}) followed by dimethyl 
ether, CH$_{3}$OCH$_{3}$, (Snyder et al. \cite{Snyder}) and
ethanol, CH$_{3}$CH$_{2}$OH, (herein denoted C$_{2}$H$_{5}$OH) 
(Zuckerman et al. \cite{Zuckerman}).
A variety of other isomers, isotopic variants and similar compounds have also been seen.
Charnley et al. (\cite{Charnley95}) 
review the distribution of complex molecules
while Ikeda et al. (\cite{Ikeda01} \& \cite{Ikeda02}) present more recent observations.
CH$_{3}$OH is seen in cold dark clouds with a relative abundance of 
10$^{-9}$ (with respect to H$_{2}$) 
while in hot cores this can be as large as 
10$^{-6}$. The highest abundances are observed in regions where grain 
mantles are likely to have recently evaporated, strongly suggesting 
mantle processing either directly or in the formation of precursors.

A variety of computer models have been constructed to investigate
molecule formation. Predominantly these are based on 
chemical reaction networks and use a \lq\lq flat\rq\rq\ (fixed) parameter space.  
As with the models presented here most are \lq\lq single point\rq\rq,
where a single representative point in a cloud is modelled. 
This is satisfactory as chemistry at another point with different 
physical parameters could be modelled simply by changing 
the parameters.
Over time the size and complexity of the reaction networks has
increased as more reaction rate data is published in the literature
and greater computer power becomes available to process more 
complex networks. 
(See Millar (\cite{Millar90}) for a review of model development.)
Millar et al. (\cite{Millar91b}) model the production of CH$_{3}$OH
by a fixed parameter gas-phase-only model 
and are unable to produce abundances in excess
of 10$^{-7}$ lending further weight to the involvement of grain mantle
processing. 
C$_{2}$H$_{5}$OH is seen in fewer sources than 
CH$_{3}$OH and all known sources are dense, hot core star-forming regions,
exactly the places where grain mantles are likely to evaporate.
Abundances are in the range 10$^{-9}$-10$^{-8}$ (with respect to H$_{2}$).
CH$_{3}$OCH$_{3}$ is often seen in the same sources as C$_{2}$H$_{5}$OH
and has recently been detected in  hot corinos
(Ceccarelli et al. \cite{Ceccarelli04}; Bottinelli et al. \cite{Bottinelli04}).
It has an abundance range of 10$^{-8}$-10$^{-6}$
and the CH$_{3}$OCH$_{3}$/C$_{2}$H$_{5}$OH abundance ratio differs 
greatly between apparently similar sources.
(Abundances with respect to H$_{2}$ 
for selected sources are listed
in Table \ref{Observed fractional abundances of complex molecules}.)

\begin{table}[ht]
\caption{Observed fractional abundances of complex molecules}
\label{Observed fractional abundances of complex molecules}
\centering
\begin{tabular}
{lrrr} 
\hline
Source - Hot Core            & CH$_3$OH             & C$_2$H$_5$OH       & CH$_3$OCH$_3$ \\
\hline
\object{NGC6334F}            & $2.0\times10^{-7}$ & $9.0\times10^{-9}$ & $4.0\times10^{-8}$ \\
\object{G327.3-0.6}          & $1.0\times10^{-7}$ & $1.0\times10^{-8}$ & $3.0\times10^{-8}$ \\
\object{G31.41+0.31}         & $9.0\times10^{-8}$ & $2.0\times10^{-8}$ & $2.0\times10^{-8}$ \\
\object{G34.3+0.2}           & $9.0\times10^{-8}$ & $6.0\times10^{-9}$ & $1.0\times10^{-8}$ \\
\object{G10.47+0.03}         & $2.0\times10^{-7}$ & $1.0\times10^{-8}$ & $3.0\times10^{-8}$ \\
\object{Sgr B2 (N)}          & $2.0\times10^{-8}$ & $1.0\times10^{-9}$ & $7.0\times10^{-10}$ \\
\object{DR21(OH)}            & $1.0\times10^{-8}$ &                   & \\
\object{W3(H$_2$O)}          & $4.0\times10^{-8}$ &                   & \\
\object{W51 e1/e2}           & $3.0\times10^{-7}$ & $9.0\times10^{-9}$ & \\
\object{Orion Compact Ridge} & $2.0\times10^{-7}$ &                   & \\
\object{Orion Hot Core}      & $1.0\times10^{-6}$ & $2.0\times10^{-8}$ & \\
\hline
Source - Hot Corino \\
\hline
\object{IRAS16293-2422}      & $1.0\times10^{-7}$ &                   & $2.4\times10^{-7}$ \\
\object{NGC1333-IRAS4A}      & $<1.0\times10^{-8}$&                   & $<2.8\times10^{-8}$ \\
\object{NGC1333-IRAS4B}      & $7.0\times10^{-7}$ &                   & $<1.2\times10^{-6}$ \\
\object{NGC1333-IRAS2A}      & $3.0\times10^{-7}$ &                   & $3.0\times10^{-8}$ \\
\hline
\end{tabular}
Data from Ikeda et al. \cite{Ikeda01}; Ikeda et al.
\cite{Ikeda02}; Bottinelli et al. \cite{Bottinelli07}
\end{table}

\pagebreak

Even under very 
favorable conditions fixed parameter gas-phase-only chemical models 
produce maximum abundances for C$_2$H$_5$OH and 
CH$_3$OCH$_3$ of the order of 10$^{-11}$
(Herbst \& Leung \cite{Herbst}; Millar et al. \cite{Millar91a}; 
Charnley et al. \cite{Charnley92}),
between two and five orders of magnitude below that observed.
Again this suggests a scenario in which mantles form on dust grains 
at some point in a cloud's lifetime when conditions are favorable.
The mantles are active and complex molecules (or their precursors) 
form on them more efficiently than in the gas phase. 
Later the cloud evolves becoming warmer and denser and the contents 
of the mantles pass back into the gas phase.
Several other gas-phase-only
models (for example Charnley et al. \cite{Charnley92};
Caselli et al. \cite{Caselli})
have demonstrated that the chemistry in hot cores is active, 
with particular gas phase reaction channels
initiated with simple species thought to originate from mantles,
a possible explanation for
observed complex molecule abundances.

A number of more in-depth models have included surface reaction chemistry. 
One of the first has been developed by 
Hasegawa et al. (\cite{Hasegawa92}) and Hasegawa \& Herbst (\cite{Hasegawa93}).
This model has 274 chemical species with 2928 reactions and  
includes 1\% (by mass) dust grains in the gas cloud. 
Molecules can freeze out onto grains 
and later pass back into the gas phase.
Once on a grain heavier species molecules are held stationary at lattice binding sites 
while lighter species, most commonly hydrogen atoms, can migrate around the grain.
When mobile light species encounter fixed heavy species the two can
combine to produce progressively larger molecules. 
While large molecules can thaw off the grain surface
there are a number of other possible desorbtion mechanisms that can act as well.
(See Section \ref{Desorption mechanisms}.) 
One additional benefit of this approach is that grain surface catalysis of 
hydrogen molecules can be modelled directly.

Shalabiea \& Greenberg (\cite{Shalabiea95b}) then take the next logical step in
interstellar cloud modelling by including variations in physical parameters.
They use a gas phase reaction rate network and interchange between gas and surface
similar to Hasegawa et al. (\cite{Hasegawa92}) and Hasegawa \& Herbst (\cite{Hasegawa93}).
However certain physical parameters (eg. density) can also change.
Shalabiea \& Greenberg (\cite{Shalabiea95b}) model these by including differential 
equations for them and solving for the relevant parameter(s) at each point 
in time in much the same way as species abundances are solved for. 
They also coin the terms \lq\lq pseudo-time-dependent\rq\rq,
where the model chemistry evolves over time while the parameters are static  
and \lq\lq real-time-dependent\rq\rq\ where both chemistry and physical parameters evolve.
Their paper provides comparison between the two types, with the gas and grain model 
including 218 chemical species and 2075 reactions.
With the more recent identification of \lq\lq hot corinos\rq\rq\
(Table \ref{Observed fractional abundances of complex molecules} \&
Section \ref{Calculating parameters}) 
where the temperature range is optimal for the grain surface production
of larger molecules and their subsequent return to the gas phase
some specific modelling of these type of objects has been done.
Garrod \& Herbst (\cite{Garrod06}) use a gas and grain reaction network 
with 655 species and 6509 reactions. The same model was used by Aikawa et al. 
(\cite{Aikawa08}).
Subsequently Garrod et al. (\cite{Garrod08}) produced an extended 
and more generalized version.
Comparison of the output of these models with the results presented here 
is made in Section \ref{Conclusion}.

As is discussed further in Section \ref{Desorption mechanisms} 
some mechanism(s) must be returning material from grain mantles
back to the gas phase since the alternative is depletion of 
molecular species heavier than H$_2$ in a timescale shorter than the
lifetime of known clouds, which is not seen.
More recent modelling work has focused on the investigating the effectiveness 
of several proposed desorption mechanisms.
Willacy \& Millar (\cite{Willacy98}) provide a set of gas and surface models
which include different desorption mechanisms and compare their effectiveness.
These models use 282 species and 4864 reactions. 
Garrod et al. (\cite{Garrod07}) specifically model formation of 
CH$_{3}$OH in a quiescent cloud and point out that if grain surface
processes are invoked to explain the abundance of CH$_{3}$OH then 
at least one non-thermal desorbtion mechanism must be active.
Tielens \& Charnley (\cite{Tielens97})
have argued that, at least in some cases, the reaction network model
cannot be justified and that a Monte Carlo simulation approach is
more appropriate. This has led to some debate in the current literature 
about the relative merits of the two methods. 
Willacy \& Millar (\cite{Willacy98}) discuss this and conclude that
currently there are computational impracticalities in using the
Monte Carlo approach in real-time-dependent models, and further that 
the two different approaches may give fairly similar results 
for systems of increasing complexity. 
A direct comparison of the two approaches by Garrod et al. 
(\cite{Garrod09}) demonstrates considerable similarity between them.
Certainly previous gas phase 
only reaction rate models have yielded results comparable with 
observations for many species.

The model used here builds on its predecessors.
Termed the \lq\lq Gas/Surface\rq\rq\ model it is a single point, 
gas and surface phase, chemical reaction network model 
with 279 species and 2968 reactions.
It includes the grain surface mechainsm used by 
Hasegawa et al. (\cite{Hasegawa92}) and Hasegawa \& Herbst (\cite{Hasegawa93})
as well as the desorption mechanisms discussed by 
Willacy \& Millar (\cite{Willacy98}).
It is also fully \lq\lq real-time-dependent\rq\rq. 
Instead of the differential equation approach taken by 
Shalabiea \& Greenberg (\cite{Shalabiea95b})
each separate time step has its own set of physical parameters
which are fed into the model from a storage file at the beginning
of each time step.
This approach has several advantages. Firstly it reduces computational
strain on the model since there is no additional equation solving
necessary. This is particularly important in situations where parameters
are changing very quickly and there are major differences between 
adjacent time steps. (For example in the final stages of a cloud collapse
where density increases drastically in a short time.)
Secondly, this approach allows the same set of software to be used to
model multiple different situations with no reprogramming at all.
Only the input parameter file has to be changed.
This allows great flexibility in the scope of scenarios that could 
be investigated. Further, in certain situations where there are
major, abrupt changes in physical parameters (eg. shocks),
the differential equation approach is highly prone to breakdown  
as generating a numerical solution for a given parameter requires some level of 
continuity between one time step and the next, wheras the parameter file 
approach avoids this problem completely.

The Gas/Surface model is used here to investigate the scenario
of complex molecule formation in grain mantles and their subsequent 
release back into the gas phase.
We consider clouds at a variety of initial densities
with different exposure to photoionization.
As these clouds collapse to become denser and darker,
their chemistry becomes more complex and their 
temperature drops allowing grain mantles to form.
Simple gas-phase dominated chemistry at the beginning leads to the 
deposition of a variety of basic molecules onto an initially bare
grain surface. Once on the grain, surface processes allow the build
up of complex molecules in a frozen mantle.
Later, as the cloud heats up, 
the mantle evaporates introducing 
the complex molecules into the gas phase with a variety of 
chemical consequences significantly different from gas phase 
only processing. 


\section{The Gas/Surface Model}
\label{The Gas/Surface Model}

\subsection{Design of the model}
\label{Design of the model}

The Gas/Surface model is used here to model a collapse situation. 
A cloud collapse begins with the cloud optically
thin and open to heating by external photons.
Self shielding of H$_2$ is treated by the method of Wagenblast
(\cite{Wagenblast}). 
As it contracts it eventually becomes optically thick 
to external photons and its inner regions cool. 
It is this cooling which allows molecules to freeze out 
onto dust grain surfaces and form mantles. Eventually continuing collapse 
of the cloud, assisted possibly by the ignition of a proto-star at
its center, causes the temperature to rise again desorping the mantles
off the grain and returning mantle processed species to the gas phase.
The model allows interchange between gas and surface phases, both freezing, 
thermal desorption and a variety of other continously acting desorbtion mechanisms,
as described by Willacy (\cite{Willacy93}) and Willacy \& Millar (\cite{Willacy98}).
(See Section \ref{Desorption mechanisms}.)
Once on the grain surface chemistry takes place as described in Section
\ref{Surface chemistry mechanism}.

The reaction networks are solved numerically using the GEAR package
(Hindmarsh \cite{Hindmarsh74}, \cite{Hindmarsh83}). 
It is the combination of both a time variable
parameter space and interchange between gas and surface phase chemistry
that makes the Gas/Surface model appropriate to investigate
complex molecule formation. A fuller description of the model design 
can be found in Hall (\cite{Hall}).

\subsection{Surface chemistry mechanism}
\label{Surface chemistry mechanism}

The surface chemistry mechanism used in the Gas/Surface Model
is taken from  Hasegawa et al. (\cite{Hasegawa92}) and 
Hasegawa \& Herbst (\cite{Hasegawa93}) with additional complex 
molecule formation reactions listed in the Appendix.
Surface species are prefixed \lq\lq $*$\rq\rq\ to distinguish 
them from their gas phase counterparts.
This scheme allows fourteen low molecular mass species
(Table \ref{Mobile surface species}) 
to be mobile on the
grain surface while all other surface species are stationary,
held at \lq\lq sites\rq\rq\ on the surface.
The adsorption energy $E_D$ (in K) is the energy needed to liberate the
species from the grain surface back into the gas phase.
Parameter values (and their references) are listed in Table 
\ref{Gas/Surface model - fixed parameters}.

The mobile species migrate around the grain surface
by \lq\lq hopping\rq\rq\ from site to site.
A mobile species arriving at an occupied site
can react with either a stationary species already there or another
mobile species which has also just arrived.
Mobile species are regarded as occupying any given site for a finite
period of time, generally about 10$^{-12}$s, the exact time being determined
by species mass and prevailing temperature.
There is an energy potential barrier separating 
adjacent sites and mobile species must possess sufficient energy to 
overcome it. For most of the mobile species \lq\lq classical\rq\rq\
type motions only are permitted. The species must have sufficient 
classical kinetic energy to cross potential barriers.
For atomic and molecular hydrogen however quantum tunnelling through 
barriers is permitted.
Which reactions take place is governed 
by a reaction set (Section \ref{Reaction set})
analogous to those used for gas 
phase reactions, each with an accompanying rate coefficient.
The grain itself is assumed to be inert and all surface 
chemistry involves the mantle only.
The classical mobile species migrate over the grain surface.
The time to move
between two adjacent surface sites (\lq\lq hopping time\rq\rq) 
$t_{hop}$ is given by :

\begin{equation}
\frac{1}{t_{hop}} = \nu_{0} {\rm exp} 
\left( \frac{ - E_b}{kT} \right)
\end{equation}

\noindent
where $T$ is the grain surface temperature
and $E_b$ is the potential
energy barrier between adjacent surface sites,
$E_b \simeq 0.3 E_D$.
The parameter $\nu_{0}$ is the characteristic vibration frequency
for the mobile species (assumed isotropic). 
Its value (in Hz) is given by :

\begin{equation}
\nu_{0} = \sqrt {\frac{2n_s E_D}{\pi^2m}}
\end{equation}

\begin{table}
\caption{Mobile surface species}
\label{Mobile surface species}
\begin{center}
\begin{tabular}
{llrl} 
\hline
     & Species    &    Adsorption         & Ref. \\
     &            &    Energy $E_D/k$ (K) &      \\
\hline
 1   &  $*$H      &    350.0   & 1 \\                  
 2   &  $*$H$_2$  &    450.0   & 1 \\                  
 3   &  $*$He     &    100.0   & 2 \\                 
 4   &  $*$C      &    800.0   & 1 \\                  
 5   &  $*$N      &    800.0   & 1 \\                  
 6   &  $*$O      &    800.0   & 1 \\                  
 7   &  $*$S      &   1100.0   & 1 \\                  
 8   &  $*$CH     &    645.0   & 3 \\                  
 9   &  $*$NH     &    604.0   & 3 \\                  
 10  &  $*$OH     &   1259.0   & 3 \\
 11  &  $*$CH$_2$ &    956.0   & 3 \\         
 12  &  $*$NH$_2$ &    856.0   & 3 \\        
 13  &  $*$CH$_3$ &   1158.0   & 3 \\         
 14  &  $*$NH$_3$ &   1107.0   & 3 \\         
\hline
\end{tabular}
\newline
\textbf{References}. 
(1) Tielens \& Allamandola \cite{Tielens87};
(2) Tielens \& Hagen       \cite{Tielens82};
(3) Allen   \& Robinson    \cite{Allen};
\end{center}
\end{table}

\noindent
where $n_s$ is the surface density of sites 
($\simeq 1.5 \times 10^{19}$ m$^{-2}$)
and $m$ is the mass of the mobile species (in kg).
This gives values of $\nu_{0}$ in the range 
10$^{12}$-10$^{13}$ s$^{-1}$. 
Hydrogen atoms and molecules are held to be quantum particles and 
their hopping times are given by the time for them to quantum tunnel
through the potential barrier :

\begin{equation}
\frac{1}{t_{hop}} = \nu_{0} {\rm exp} 
\left[ 
\frac{ - 2a}{\hbar}
\sqrt{2mE_b} 
\right]
\end{equation}

\noindent
where $a$ is the separation between two adjacent sites 
($\simeq 10^{-10}$ m).
The diffusion time, $t_{diff}$, for a mobile 
species to 
scan the entire grain surface is given by :

\begin{equation}
t_{diff} = N_st_{hop}
\end{equation}

\noindent
where $N_s$ is the total number of surface sites on 
a grain ($N_s$ = 10$^{6}$).
The surface reaction rate coefficient $k_{ij}$ 
between two surface species $i$ and $j$ by classical 
diffusion is given by :

\begin{equation}
k_{ij} = 
\frac{\kappa_{ij} \left(\frac{1}{t_i} + \frac{1}{t_j}\right)}{N_sn_g}
\end{equation}

\noindent
where $t_i$ and $t_j$ are the hopping times for species 
$i$ and $j$ and $n_g$ is the number density of grains. 
A gas:grain ratio of 100:1 by mass, 1:1.33 $\times$ 10$^{-12}$ by number
is used.
The factor $\kappa_{ij}$ is the reaction probability for a specific reaction.
$\kappa_{ij}$ = 1.0 unless the associated activation
energy ($E_a$) is non-zero.
Non-zero activation energy enables certain reaction channels 
to be favoured over others. 
In the classical case :
\\
\begin{equation}
\kappa_{ij} = 
{\rm exp} \left(
\frac{-E_a}{kT}
\right) 
\end{equation}

\noindent
In the quantum case where at least one of the reactants is a hydrogen
atom or molecule :
\\
\begin{equation}
\kappa_{ij} = 
{\rm exp} \left[\frac{-2a}{\hbar} \sqrt{2\mu E_a}\right]
\end{equation}

\noindent
where $\mu$ (kg) is the reduced mass
in a two-body collision.

\subsection{Desorption mechanisms}
\label{Desorption mechanisms}

Desorption mechanisms can be divided into two basic classes,
continuously acting which remove a few 
atoms/molecules at any one time and irregular 
such as shocks, which occur at random intervals but return 
most or all of a grain mantle to the gas phase. 
In the scenarios modelled here there are no shocks 
or similar mechanisms of sufficient 
strength to desorb mantles and continous mechanisms dominate.
Willacy (\cite{Willacy93}) provides an in-depth treatment of continuous
mechanisms and four of them have been included in the 
models presented here :

\begin{enumerate}
\item Mantle Explosion      
\item Direct UV Photodesorption 
\item Cosmic Ray Induced Photodesorption
\item Direct Heating by Cosmic Rays  
\end{enumerate}

For each desorption mechanism an equation is provided 
in the reaction set with its own rate coefficient $k$ :

\begin{equation}
{\rm Surface\ species}\ +\ {\rm desorption\ mechanism}\
\rightarrow\ {\rm gas\ species}
\end{equation}

One of these four mechanisms, direct UV photodesorption, 
dominates when a cloud is in the early stages of
collapse and still optically thin. Once collapse renders 
the bulk of the cloud opaque to external photons its effects 
are negligible and the other three mechanisms are more significant.
However in the final stages of collapse direct thermal desorption of 
grain mantles takes place and this outstrips all four
continuous desorption mechanisms.
The rate of evaporation is given by 
(Hasegawa et al. \cite{Hasegawa92}) :
\begin{equation}
k_{evap} = \nu_{0} {\rm exp} \left( \frac{-E_D}{kT} \right)
\end{equation}
\noindent
For those species that are not mobile on the surface default values are 
$\nu_{0}$ = 10$^{12} {\rm s}^{-1}$ and 
$E_b = 2,000 {\rm K}$.

\subsection{Reaction set}
\label{Reaction set}

The basic gas phase reactions are chosen to give a representative
model of an interstellar cloud. Beginning with the UMIST astrochemical
reaction rates database, RATE95 (Millar et al. \cite{Millar97}), 
all reactions involving chlorine, phosphorus, iron, silicon, sodium 
and sulphur are removed along with those involving species with five or
more carbon atoms. This yields a reaction set based on hydrogen, helium,
carbon, oxygen and nitrogen. Magnesium reactions are included to 
ensure a \lq\lq token metal\rq\rq\ and source of electrons.
Additional provision is made for species formed only on
grain surfaces which can pass into the gas phase.
An extra set of reactions allowing gas
phase destruction of these species has been added to the 
gas phase reaction set,
listed in the Appendix
(Table \ref{Gas phase reactions used for complex molecule destruction}).
 
The surface phase reaction set (Table
\ref{Solid phase reactions used for complex molecule formation})
is constructed with reference to 
those reactions provided by Hasegawa et al. (\cite{Hasegawa92}) and
Hasegawa \& Herbst (\cite{Hasegawa93}).
However it is very much optimized to allow the 
production of $*$CH$_3$OH,
$*$C$_2$H$_5$OH and $*$CH$_3$OCH$_3$.
The reactions listed by Hasegawa et al.
contain a number of laboratory 
measured potential energy barriers which slow down 
the rate of formation of particular species.
Only a few such barriers are used here. 
Further, the reaction set includes almost
entirely reactions orientated to the formation of 
complex molecules.
More reactions to produce other
species would reduce the
complex molecule abundance by draining  
\lq\lq raw material\rq\rq.
It is recognized that this orientation of the reaction set
is a limitation of the Gas/Surface model.
However, it is regarded as a valid test of the hypothesis
of complex molecule formation on grain surfaces.
If, given all these advantages, such molecules still could not
be produced by grain surface catalysis then the original
supposition of such a production method would have to come 
under deeper scrutiny.
 
A small number of surface reactions are included that are not 
directly involved in complex molecule formation.
These are included because they are are believed to be 
significant to overall surface chemistry.
The most basic of all is the grain formation of H$_2$,
whose equation is :

\begin{equation}
{\rm *H}\ +\ {\rm *H}\ \rightarrow\ {\rm H}_2 
\end{equation}

It is considered that the H$_2$ molecule produced passes directly
into the gas phase without the need for any specific desorption
process. This is consistent with quantum calculations
which indicate the resultant molecule is formed with sufficient 
energy to eject it directly from the surface.
Other reactions form small molecules
which are frequently seen in high abundances in regions where
grain mantles are thought to have recently evaporated.
The basis of nitrogen chemistry is formation of
$*$NH$_3$ by the route :

\begin{equation}
{\rm *N}\ +\ {\rm *H}\ \rightarrow\ {\rm *NH}\ +\ {\rm *H}
\rightarrow\ {\rm *NH_2} +\ {\rm *H} \rightarrow\ {\rm *NH_3}
\end{equation}

Direct combination of atoms can also form $*$CN, $*$NO,
$*$HNO, $*$HCN and $*$HNC.
$*$O$_2$ and $*$OH can also form directly and $*$OH can further 
react to form $*$H$_2$O.
Other combination reactions allow the formation
of $*$CO and $*$CH$_4$.
$*$CO and $*$H can also react to form $*$HCO and $*$H$_2$ :

\begin{equation}
{\rm *CO}\ +\ {\rm *H}\ \rightarrow\ 
{\rm *HCO}\ +\ {\rm *H}\ \rightarrow\ 
{\rm *CO} +\ {\rm *H_2} 
\end{equation}

$*$H$_2$CO can be formed and destroyed by the reactions :

\begin{equation}
{\rm *CH_2}\ +\ {\rm *O}\ \rightarrow\ {\rm *H_2CO}
\end{equation}

\begin{equation}
{\rm *H_2CO}\ +\ {\rm *H}\ \rightarrow\ {\rm *HCO} +\ {\rm *H_2} 
\end{equation}

There are two possible routes to $*$CO$_2$, which is 
stable on the grain surface and does not react further :

\begin{equation}
{\rm *CO}\ +\ {\rm *OH}\ \rightarrow\ {\rm *CO}_2 +\ {\rm *H} 
\label{co2-1}
\end{equation}

\begin{equation}
{\rm *HCO}\ +\ {\rm *O}\ \rightarrow\ {\rm *CO}_2 +\ {\rm *H} 
\label{co2-2}
\end{equation}

Direct combination of $*$CO with $*$O has been omitted as it
is believed that this reaction is extremely inefficient
(Grim \& d'Hendecourt \cite{Grim}; Breukers \cite{Breukers}).
Direct combination of carbon atoms to form $*$C$_2$ has also
been omitted to simplify the surface chemistry.
If it occurred it would lead by hydrogenation to 
the presence of significant amounts of $*$C$_2$H$_6$,
not currently detected. C$_2$H$_6$ is
seen in the cometry coma of comets Hyakutake and Hale-Bopp
(Mumma et al. \cite{Mumma})
and it is not known if this formed on the surface or in the gas phase.
The bulk of surface reactions comprise
routes to the complex molecules $*$CH$_3$OH, $*$C$_2$H$_5$OH and 
$*$CH$_3$OCH$_3$. 
$*$CH$_3$OH is formed by combining $*$CH$_{\rm n}$ and $*$OH 
(n = 0-3) :

\begin{equation}
{\rm *CH_n}\ +\ {\rm *OH}\ \rightarrow\ {\rm *CH_nOH}
\end{equation}

The resultant is then hydrogenated as necessary to 
form $*$CH$_3$OH. Alternative routes are provided by :

\begin{equation}
{\rm *CH_3}\ +\ {\rm *O}\ \rightarrow\ {\rm *CH_3O}
+\ {\rm *H}\ \rightarrow\ {\rm *CH_3OH}
\end{equation}

\begin{equation}
{\rm *CO}\ +\ {\rm *H}\ \rightarrow\ {\rm *COH}
\end{equation}

$*$COH can then hydrogenate. Reaction between $*$CO and $*$H
can also produce $*$HCO and the ratio of the two reactions
is controlled by activation energy barriers.
Routes to $*$C$_2$H$_5$OH and $*$CH$_3$OCH$_3$ involve
forming a \lq\lq spine\rq\rq\ of C-C-O or C-O-C and then 
hydrogenating until the molecule saturates.
This is accomplished by the reactions :

\begin{equation}
{\rm *CO}\ +\ {\rm *C}\ \rightarrow\ {\rm *CCO}
\label{cco}
\end{equation}

\begin{equation}
{\rm *CO}\ +\ {\rm *C}\ \rightarrow\ {\rm *COC}
\label{coc}
\end{equation}

Activation energy barriers can be included 
which slow down particular reactions.
Some are from laboratory measurements.
(See Hasegawa et al. (\cite{Hasegawa92}) and references therein.)
Those for the production of 
${\rm *CO}_2$ (eqns. \ref{co2-1} \& \ref{co2-2} above),
${\rm *CCO}$ (eqn. \ref{cco} above) and ${\rm *COC}$ (eqn. \ref{coc} above)
have been set to prevent major overabundances of these species.
A full list of all surface reactions,
with applicable activation energies, is given in the Appendix
(Table
\ref{Solid phase reactions used for complex molecule formation}).
It is recognized that the surface reaction set
used is an approximation. 
It is constructed by interpolating between measured surface reactions,
known or suspected processes in surface chemistry and the aimed-for
result of producing complex molecule abundances and a grain mantle 
composition compatible with observations.

\subsection{Species set and starting abundances}
\label{Species set and starting abundances}

The species set is derived from the reaction set.
(Full listing in the Appendix, Tables
\ref{Gas/Surface model - surface phase species set} and
\ref{Gas/Surface model - gas phase species set}.)
All species can exist in the gas phase while
only neutral species exist on the grain surface.
Any ion freezing out is assumed to be neutralized. 
For ions with no neutral equivalent it is assumed they
break up in exactly the same way as by electron 
neutralization in the gas phase.
The starting abundances (Table
\ref{Gas/Surface model - starting abundances})
are from Hasegawa et al. (\cite{Hasegawa92}).
Originally used by Leung et al. (\cite{Leung})
they are believed to be a reasonably accurate 
representation of the abundances found in an interstellar cloud,
particularly one liable to collapse into a hot core or hot corino.
Earlier gas phase only models derived using them are consistent
with observations within the known accuracy of the chemistry. 

\begin{table}[h]
\caption{Gas/Surface model - starting abundances}
\label{Gas/Surface model - starting abundances}
\centering
\begin{tabular}
{ll} 
\hline
Species     & Abundance w.r.t. (H + H$_2$) \\
\hline
H           & $1.00$ \\
He          & $1.40 \times 10^{-1}$ \\
C           & $7.30 \times 10^{-5}$ \\
O           & $1.76 \times 10^{-4}$ \\
N           & $2.14 \times 10^{-5}$ \\
Mg          & $7.00 \times 10^{-9}$ \\
\hline
\end{tabular}
\end{table}


\section{Model Parameters}
\label{Model Parameters}

\subsection{Calculating parameters}
\label{Calculating parameters}

\begin{table*}[ht]
\caption{Gas/Surface model - fixed parameters}
\label{Gas/Surface model - fixed parameters}
\centering
\begin{tabular}
{lllll} 
\hline
Parameter                                          & Value                  & Ref. \\
\hline
UV Radiation field                                 & 3.0$\times$10$^{8}$ cm$^{-2}$s$^{-1}$Hz$^{-1}$ & 1 \\
Cosmic ray ionization rate                         & 1.3$\times$10$^{-17}$s$^{-1}$                  & 2 \\
Grain albedo                                       & 0.5                                            & 3 \\ 
Sticking coefficient                               & 1.0                                            & 4 \\   
Sticking coefficient for hydrogen atoms            & 0.3                                            & 5 \\
Yield per UV photon on grain mantle impact         & 10$^{-6}$                                      & 6 \\
Surface density of sites                           & 1.5$\times$ 10$^{15}$ cm$^{-2}$                & 4 \\
Number of sites per grain                          & 10$^{6}$                                       & 4 \\             
Fractional density of grains (w.r.t. Density of H) & 1.33$\times$10$^{-12}$                         & 4 \\
Average grain radius                               & 10$^{-5}$ cm                                   & 4 \\
Barrier width between adjacent sites               & 10$^{-8}$ cm                                   & 4 \\
Grain velocity (non-thermal)                       & 10$^{4}$ cms$^{-1}$                            & 7 \\             
\hline
\end{tabular}
\newline
\textbf{References}. 
(1) Draine                          \cite{Draine};
(2) van Dishoeck \& Black           \cite{Dishoeck};
(3) Whittet                         \cite{Whittet92}; 
(4) Hasegawa et al.                 \cite{Hasegawa92};   
(5) Brown                           \cite{Brown};
(6) Willacy                         \cite{Willacy93};
(7) V$\stackrel{..}{\rm o}$lk et al.\cite{Volk};  
\end{table*}

As used here only four physical parameters are varied 
to model cloud collapse : 
density, visual extinction, gas temperature and grain temperature. 
All other parameters can be regarded as fixed and their
values are given in Table 
\ref{Gas/Surface model - fixed parameters}.
The definitive factor is density, which controls the speed of collapse.
In all cases here the cloud is assumed to be spherical and of uniform density.
This is recognized as an approximation,
possible effects of non-uniform density in star forming clouds
are discussed by Shu et al. (\cite{Shu}). 
For the cloud to begin to contract under its own self-gravity
its mass must exceed the Jeans mass ($M_J$) for its density.
For the cloud densities modelled, the masses and corresponding 
Jeans mass are shown in 
Table 
\ref{Gas/Surface model - cloud parameters},
which also gives the relevant free-fall times ($t_{ff}$) and radii.

\begin{table*}[ht]
\caption{Gas/Surface model - cloud parameters}
\label{Gas/Surface model - cloud parameters}
\centering
\begin{tabular}
{llllllll} 
\hline

Initial Number      & Mass        & $M_J$       & t$_{ff}$            & Initial Radius & Final Radius & Initial    & Initial \\ 
Density (cm$^{-3}$)  & ($M_\odot$)  & ($M_\odot$)  & (years)            & (parsecs)      & (parsecs)    & A$_v$ (Mag) & Temp. (K)\\
\hline
100                 & 1700        & 1692.7     & 4.40$\times$10$^6$  & 5.0818         & 0.065        & 1.39       & 48.5 \\
1,000               &  150        &  144.3     & 1.39$\times$10$^6$  & 1.0501         & 0.034        & 2.86       & 20.2 \\
10,000              &   26        &  25.97     & 4.40$\times$10$^5$  & 0.2717         & 0.006        & 7.41       & 13.9 \\
\hline
\end{tabular}
\end{table*}

The cloud collapse is halted at a density of 10$^8$ cm$^{-3}$, 
observations of hot cores and hot corinos yield densities 
10$^{6}$-10$^{9}$ cm$^{-3}$.
As the data in the Appendix demonstrate,
by the end of the collapse all the hydrogen in the cloud 
has become molecular.
(H$_2$ abundance = 5.0 $\times 10^{-1}$ with respect to total hydrogen.)
$A_V$ is calculated from Brown (\cite{Brown})
and its increase is stopped at 60 magnitudes
as further increases in have negligible 
effect on the chemistry.

All cloud collapse models presented here assume that 
gas and grain temperature are the same.
Since the gas and grains are closely coupled and the temperature 
mainly varies as a smooth function it is unlikely
that they would be significantly different over the temperature region where
the grains have mantles.
As a cloud collapses gravitational energy liberated 
can be efficiently radiated in the infrared and submillimeter
bands (Dyson \& Williams \cite{Dyson}).
However, at some point the temperature must rise
to begin nuclear processing.
It is assumed that below a density of 10$^6$ cm$^{-3}$
the cloud can efficiently radiate and 
temperature ($T$) in this region is found from a curve fitting 
technique derived by Tarafdar et al. (\cite{Tarafdar}) :

\begin{equation}
\label{eq:tempbelow10^6}
T = \frac{163}
{
2.5 + log_e (n) - log_e (1 + 500 exp(-1.8A_V))
}
\end{equation}
where $n$ is number density and $A_V$ visual extinction.
Above $n = 10^6$ cm$^{-3}$ the temperature is 
determined by the release of gravitational energy :

\begin{equation}
\label{eq:tempabove10^6}
T = T_\circ
\left(
\frac{n}{n_\circ}
\right)
^\frac{1}{3}
\end{equation}
where $T_\circ$ is the temperature at 10$^6$ cm$^{-3}$.
From Tarafdar et al. $T_\circ = 10$ K and n$_\circ$ = 10$^6$ cm$^{-3}$.
The final temperature obtained is $T$ = 46.4 K at the maximum number
density $n$ = 10$^8$ cm$^{-3}$. 
The final environments produced by the models have a temperature 
of $T$ = 46.4 K, a number density $n$ = 10$^8$ cm$^{-3}$,
an $A_V$ of 60 mags and all remaining hydrogen molecular.

Recently a number of \lq\lq hot corinos\rq\rq\
(Ceccarelli et al. \cite{Ceccarelli04}, Bottinelli et al. \cite{Bottinelli04})
have been investigated.
These are regions of low mass ($\simeq$1 M$_\odot$) star formation
embedded within and condensed from molecular clouds.
They have typical sizes of $\leq$200 AU, temperatures of $\simeq$100 K
and densities of $\geq$10$^{7}$ cm$^{-3}$. 
(Cazaux et al. \cite{Cazaux}, Bottinelli et al. \cite{Bottinelli04} and references within.)
Complex molecules have been detected in four hot corino sources, listed in Table
\ref{Observed fractional abundances of complex molecules}.

\subsection{Collapse scenarios}
\label{Collapse scenarios}

\begin{table*}
\caption{Collapse scenario model parameters}
\label{Collapse scenario model parameters}
\centering
\begin{tabular}
{llllll} 
\hline
                      & Initial Number      & Parameter           & Minimum            & Plateau \\
                      & Density (cm$^{-3}$)  &                     & Temperature        & Point   \\
\hline
Model 1               & 100                 & Time (yrs)          & $4.36 \times 10^6$  & $4.39 \times 10^6$  \\ 
Free-fall             & Atomic              & $n$ (cm$^{-3}$)      & $6.97 \times 10^5$  & $1.00 \times 10^8$   \\ 
\hline
Model 2               & 1,000               & Time (yrs)          & $1.36 \times 10^6$  & $1.39 \times 10^6$  \\ 
Free-fall             & Atomic              & $n$ (cm$^{-3}$)      & $6.78 \times 10^5$  & $1.00 \times 10^8$   \\ 
\hline
Model 3               & 10,000              & Time (yrs)          & $4.18 \times 10^5$  & $4.38 \times 10^5$  \\ 
Free-fall             & Atomic              & $n$ (cm$^{-3}$)      & $9.47 \times 10^5$  & $1.00 \times 10^8$   \\ 
\hline
Model 4               & 100                 & Time (yrs)          & $2.18 \times 10^7$  & $2.19 \times 10^7$  \\ 
Retarded ($\times$ 5) & Atomic              & $n$ (cm$^{-3}$)      & $1.00 \times 10^6$  & $1.00 \times 10^8$   \\  
\hline
Model 5               & 1,000               & Time (yrs)          & $6.84 \times 10^6$  & $6.93 \times 10^6$  \\ 
Retarded ($\times$ 5) & Atomic              & $n$ (cm$^{-3}$)      & $9.87 \times 10^5$  & $1.00  \times 10^8$  \\ 
\hline
Model 6               & 10,000              & Time (yrs)          & $2.07 \times 10^6$  & $2.19 \times 10^6$   \\ 
Retarded ($\times$ 5) & Atomic              & $n$(cm$^{-3}$)       & $7.73 \times 10^5$  & $1.00 \times 10^8$   \\ 
\hline
Model 7               & 10,000              & Time (yrs)          & $4.18 \times 10^5$  & $4.38 \times 10^5$  \\ 
Free-fall             & Molecular           & $n$ (cm$^{-3}$)      & $9.47 \times 10^5$  & $1.00 \times 10^8$   \\ 
\hline
Model 8               & 10,000              & Time (yrs)          & $2.07 \times 10^6$  & $2.19 \times 10^6$  \\ 
Retarded ($\times$ 5) & Molecular           & $n$(cm$^{-3}$)       & $7.73 \times 10^5$  & $1.00 \times 10^8$   \\ 
\hline
\end{tabular}
\end{table*}

In all, eight different cloud collapse scenarios are investigated,
designated models 1-8. Models 1-3 are free-fall collapses 
with starting abundances of 100, 1,000 and 10,000 cm$^{-3}$,
regarded as slow, medium and fast. 
The observed formation rate of new stars is too low
for all stars in the galaxy to be formed by free-fall collapse.
It is widely asserted that some mechanism supports clouds 
against collapse,
the most commonly suggested being magnetic fields
(Shu et al. \cite{Shu}).
For this reason models 4-6 are retarded collapses,
again with starting abundances of
100, 1,000 and 10,000 cm$^{-3}$.
No specific retardation mechanism is incorporated in the
Gas/Surface model nor is one necessary. 
Retardation is accounted for by arbitrarily slowing down the 
decrease in radius by a factor of 5.0.
This ensures that the cloud takes longer
to collapse with corresponding
further evolution in the chemistry.
Apart from retardation the parameter change
in density, gas and grain temperature and visual extinction
follows exactly the same pattern as in models 1-3.

For the purposes of direct comparison as many parameters as possible
are kept constant for all collapses.
At a density of 100 cm$^{-3}$ the hydrogen is likely to be 
almost completely atomic.
Observed clouds of this density 
are optically thin and exposure to background 
starlight ensures an ample supply of UV photons to break up H$_2$. 
At 1,000 cm$^{-3}$ this is still approximately true though
some H$_2$ does form. However at the high density
case of 10,000 cm$^{-3}$ it is unphysical. The clouds are 
sufficiently dense that they will be predominantly H$_2$.
For this reason the final models, 7 and 8, replicate the free-fall 
and retarded collapses 
from 10,000 cm$^{-3}$ with their initial hydrogen abundance
molecular instead of atomic.

Essensially the chemistry during cloud collapse can be divided into 
three approximate segments :

\begin{enumerate}
\item
Pre-freeze out.
Gas phase chemistry dominates, similar to that in gas phase only models. 

\item
Freeze out.
As the temperature falls and the visual extinction ($A_V$) increases 
(curtailing photodesorption), freeze out occurs and 
surface reactions dominate the chemistry.
This occurs around the minimum temperature, $\simeq$ 10.0 and 10.2 K.

\item
Post-freeze out. 
Once the temperature rises the entire grain mantle
passes back into the gas phase. 
Grain surface products are suddenly released into a very
dense, warm, dark cloud environment.
This leads to a rapidly changing chemistry where 
a wide variety of different species interact.
\end{enumerate}

Table \ref{Collapse scenario model parameters}
lists the parameter values for each model at the beginning
of its run, the point of minimum temperature and the \lq\lq plateau point\rq\rq.
The plateau point is where parameter change ceases. 
In all cases at the plateau point 
the density is 10$^8$ cm$^{-3}$, temperature 46.4 K 
and visual extinction 60.0 magnitudes. 
For the $2.0 \times 10^5$ years beginning at the plateau point
each model most closely represents the chemistry of a hot corino. 
All model runs terminate $2.0 \times 10^5$ years after the plateau point is reached.

\section{Discussion}
\label{Discussion}

The Appendix contains output data from each of the eight models.
The most abundant gas phase species at both the
plateau point and 2.0 $\times 10^5$ years later
are listed in Tables
\ref{Model 1 - Free fall collapse from 100 cm$^{-3}$ - 
most abundant gas phase species}
-
\ref{Model 8 - Retarded collapse from 10,000 cm$^{-3}$ -
H$_2$ starting abundance - most abundant gas phase species}.
Variation in gas phase species abundance over time is shown in Figures
\ref{Model1 - Free fall collapse from 100 cm$^{-3}$ - gas phase data}
-
\ref{Model 8 - Retarded collapse from 10,000 cm$^{-3}$ - H$_2$ starting abundance - gas phase data}.
The results are discussed in this section.

\subsection{Mantle composition}
\label{Mantle composition}

\begin{table*}[ht]
\caption{Observed interstellar ices
(See text for references)}
\label{Observed interstellar ices}
\centering
\begin{tabular}
{lrrrrrrrr} 
\hline
Source                   &   $*$H$_{2}$O     & $*$HCOOH       & $*$CH$_{3}$OH & $*$CH$_{4}$ & $*$CO$_{2}$ & $*$CO     & $*$NH$_{3}$ & $*$H$_2$CO\\ 
\hline

\object{B1-b}            &   100.00         &     3.1        &    11.2      &  3.3        &            &           &            &     \\
\object{IRAS 08242-5050} &   100.00         &     2.7        &     5.5      &  5.0        &            &           &            &     \\
\object{IRAS 15398-3359} &   100.00         &     1.9        &    10.3      &  6.0        & 35.24      &  5.63     &            &     \\
\object{SVS 4-5}         &   100.00         &     4.6        &    25.2      &  6.1        & 30.46      & 13.09     &            &     \\
\object{L1014 IRS}       &   100.00         &     5.4        &     3.1      &  7.1        &            &           &            &     \\
\object{W33A}            &   100.00         &     5.2        &    14.7      &             & 12.49      &  7.37     & 15.0       & 6.0 \\
\object{GL 2136}         &   100.00         &     5.1        &     8.5      &             & 13.24      & 10.19     &            &     \\
\object{GL 7009S}        &   100.00         &     2.5        &    31.3      &             &            &           &            &     \\
\object{NGC7538 IRS9}    &   100.00         &                &     7.5      &             &            &           & 13.0       & 4.0 \\

\hline
\end{tabular}
\end{table*}

\begin{table*}[ht]
\caption{Mantle composition w.r.t H$_2$O at maximum mantle thickness for each model}
\label{Mantle composition w.r.t H$_2$O at maximum mantle thickness for each model}
\centering
\begin{tabular}
{lrrrrrrrr}
\hline
 Species          &     Model 1 &    Model 2 &    Model 3 &     Model 4 &    Model 5 &    Model 6 &    Model 7 & Model 8 \\
\hline
 $*$H$_2$O        &      100.00 &     100.00 &     100.00 &      100.00 &     100.00 &     100.00 &     100.00 &  100.00 \\
 $*$CH$_4$        &       34.23 &      28.62 &      32.50 &       28.34 &            &      31.51 &       3.08 &    2.45 \\
 $*$NH$_3$        &        5.94 &       3.88 &       5.04 &             &            &            &       1.14 &         \\
 $*$N$_2$         &        3.75 &       5.60 &       4.39 &        7.87 &      13.76 &       6.31 &       8.46 &   10.50 \\
 $*$CO            &        3.65 &       7.15 &       3.88 &       14.73 &      69.53 &       4.42 &      55.66 &   71.91 \\
 $*$CO$_2$        &             &       1.69 &            &        7.16 &      24.46 &       2.22 &      22.12 &   28.43 \\
 $*$O$_2$         &        1.24 &            &            &        5.65 &      12.79 &       2.17 &       2.77 &   15.49 \\
 $*$HNO           &             &            &            &        1.92 &       3.71 &       1.45 &       9.16 &   11.68 \\
 $*$C$_4$H        &             &            &            &             &            &            &            &    1.17 \\

\hline
 $*$CH$_3$OH      &  7.98                & 14.70               & 11.46               &  10.32               & 12.12               & 13.05               & 1.91               & 1.57                \\
 $*$C$_2$H$_5$OH  &  $1.19\times 10^{-3}$ & $7.74\times 10^{-4}$ & $7.64\times 10^{-4}$ &  $9.50\times 10^{-4}$ & $1.28\times 10^{-3}$ & $6.21\times 10^{-4}$ & $3.49\times 10^{-2}$ & $2.43\times 10^{-2}$ \\
 $*$CH$_3$OCH$_3$ &  $5.80\times 10^{-7}$ & $2.60\times 10^{-7}$ & $2.10\times 10^{-7}$ &  $9.00\times 10^{-8}$ & $9.47\times 10^{-5}$ & $5.50\times 10^{-7}$ & $2.24\times 10^{-4}$ & $1.05\times 10^{-4}$ \\
\hline
\end{tabular}
\end{table*}

\begin{table*}[ht]
\caption{Mantle layering}
\label{Mantle layering}
\centering
\begin{tabular}
{llllll}
\hline
 & & Model 1 & Model 2 & Model 3 & Model 4\\
\hline
Maximum     & Layers              & 184               & 169               & 179               & 154               \\ 
Thickness   & Time (years)        & $4.38\times 10^6$ & $1.38\times 10^6$ & $4.28\times 10^5$ & $2.19\times 10^7$ \\ 
            & Density (cm$^{-3}$)  & $5.95\times 10^6$ & $5.02\times 10^6$ & $3.39\times 10^6$ & $5.62\times 10^6$ \\ 
            & Temperature (K)     & 18.1              & 17.1              & 15.0              & 17.8              \\
\hline
 & & Model 5 & Model 6  & Model 7   & Model 8  \\
\hline
Maximum     & Layers              & 121               & 169               & 126               & 119               \\ 
Thickness   & Time (years)        & $6.90\times 10^6$ & $2.14\times 10^6$ & $4.28\times 10^5$ & $2.14\times 10^6$ \\ 
            & Density (cm$^{-3}$) & $8.81\times 10^6$ & $3.20\times 10^6$ & $3.39\times 10^6$ & $3.20\times 10^6$ \\ 
            & Temperature (K)     & 20.7              & 14.7              & 15.0              & 14.7              \\
\hline
\end{tabular}
\end{table*}

\begin{table*}[ht]
\caption{Gas phase abundances - freeze out}
\label{Gas phase abundances - freeze out}
\centering
\begin{tabular}
{llllllll} 
\hline
                 & Time (years) & C & CO & CO$_2$ & H & H$_2$ & O \\ 
\hline
Model 1 & 3.67 $\times 10^6$   & 6.76 $\times 10^{-5}$   & 8.34 $\times 10^{-7}$  & 7.95 $\times 10^{-14}$  & 7.79 $\times 10^{-1}$   & 1.11 $\times 10^{-1}$  & 1.71 $\times 10^{-4}$ \\
Model 2 & 7.94 $\times 10^5$   & 5.81 $\times 10^{-5}$   & 1.27 $\times 10^{-5}$  & 3.97 $\times 10^{-11}$  & 7.64 $\times 10^{-1}$   & 1.18 $\times 10^{-1}$  & 1.61 $\times 10^{-4}$ \\
Model 3 & 1.25 $\times 10^4$   & 6.99 $\times 10^{-5}$   & 9.23 $\times 10^{-7}$  & 1.07 $\times 10^{-12}$  & 9.73 $\times 10^{-1}$   & 1.33 $\times 10^{-2}$  & 1.71 $\times 10^{-4}$ \\
Model 4 & 1.81 $\times 10^7$   & 5.67 $\times 10^{-5}$   & 1.13 $\times 10^{-5}$  & 1.46 $\times 10^{-11}$  & 3.10 $\times 10^{-1}$   & 3.45 $\times 10^{-1}$  & 1.60 $\times 10^{-4}$ \\
Model 5 & 4.42 $\times 10^6$   & 3.69 $\times 10^{-6}$   & 6.55 $\times 10^{-5}$  & 3.69 $\times 10^{-7}$   & 2.20 $\times 10^{-1}$   & 3.90 $\times 10^{-1}$  & 9.64 $\times 10^{-5}$ \\
Model 6 & 1.25 $\times 10^4$   & 6.99 $\times 10^{-5}$   & 9.23 $\times 10^{-7}$  & 1.07 $\times 10^{-12}$  & 9.73 $\times 10^{-1}$   & 1.33 $\times 10^{-2}$  & 1.71 $\times 10^{-4}$ \\
Model 7 & 2.51 $\times 10^4$   & 6.08 $\times 10^{-5}$   & 6.07 $\times 10^{-6}$  & 5.95 $\times 10^{-9}$   & 3.48 $\times 10^{-5}$   & 5.00 $\times 10^{-1}$  & 1.61 $\times 10^{-4}$ \\
Model 8 & 2.51 $\times 10^4$   & 6.08 $\times 10^{-5}$   & 6.06 $\times 10^{-6}$  & 5.94 $\times 10^{-9}$   & 3.47 $\times 10^{-5}$   & 5.00 $\times 10^{-1}$  & 1.61 $\times 10^{-4}$ \\
\hline
\end{tabular}
\end{table*}

The test of any theoretical model is to compare it to
observations. Table 
\ref{Observed interstellar ices}
presents observational data on interstellar ices 
towards a number of sources. The abundances shown are relative
to *H$_{2}$O and have been measured by 
Boogert et al. (\cite{Boogert08}) for $*$HCOOH \& $*$CH$_{3}$OH, 
Pontoppidan et al. (\cite{Pontoppidan}) for CO \& CO$_2$,
$\stackrel{..}{\rm O}$berg et al. (\cite{Oberg08}) for CH$_{4}$ and 
Gibb et al. (\cite{Gibb}) for $*$NH$_{3}$ \& $*$H$_2$CO.
For data taken from Boogert et al. (\cite{Boogert08})
measured values have been taken, observations 
where a lower limit only has been established have been ommitted. 
Further Boogert et al. (\cite{Boogert08}) data for $*$NH$^{+}_{4}$
has been ommitted as the authors themselves point out that the inclusion of 
certain components in the abundance estimation is still a matter of debate.

It is assumed that ice measurements reflect the composition
of grain mantles along the measured lines of sight.
No measurements are available for mantles in hot cores or hot corinos
as within these regions the temperature is sufficiently high
that any grains have been fully thermally desorbed.
Observations made of other, cooler regions are therefore
the best available indicator of grain mantle composition.
All models produce mantles with a predominance of $*$H$_2$O,
(absolute composition 40-64\%), agreeing with observations. 
Table \ref{Mantle composition w.r.t H$_2$O at maximum mantle thickness for each model}
presents mantle composition at greatest thickness as a percentage of $*$H$_2$O abundance
for all mantle species with abundance $\geq$ 1\% $*$H$_2$O 
and for the complex molecules regardless of their contribution.
Table \ref{Mantle layering} shows the conditions during each model
when the mantle is at its greatest thickness.
The point of greatest thickness is defined as that timestep with the 
largest number of surface layers. Where this is the same for two 
or more timesteps the latest is taken. 

$*$CO is well modelled in models 1-4 and 6 and overproduced in the other three.
For models 7 and 8 this is a consequence of the molecular hydrogen starting abundance.
In both cases atomic carbon and oxygen are able to preferentially combine with
each other due to the absence of atomic hydrogen which would otherwise convert 
them to $*$H$_2$O and $*$CH$_4$. Model 5 appears to be an optimum case 
(with initial atomic hydrogen) for the production of $*$CO and also subsequently 
$*$C$_2$H$_5$OH and $*$CH$_3$OCH$_3$, hence the higher proportion
of all three of these species.
Table \ref{Gas phase abundances - freeze out} bears this out, 
showing an order of magnitude lower carbon abundance at freeze out time 
in model 5 compared to the other models. 
Correspondingly both $*$CO and $*$CO$_2$ are significantly more abundant in model 5.
Both these are preferentially formed in the gas phase before freeze out.
$*$CO$_2$ is well modelled in models 4, 5, 7 and 8 and underproduced in all others.
Again in models 7 and 8 the absence of initial atomic hydrogen 
is the major factor here, while in models 4 and 5 it seems that the longer
time period before freeze out enhances $*$CO$_2$ production,
for the same reasons as it also enhances $*$CO production.
Exactly the same factors result in $*$CH$_4$ being well modelled 
in models 5, 7 and 8 and overproduced in all others. 
The faster collapses (models 1, 2, 3 and 6) freeze out atomic carbon 
before it can form molecules in the gas phase and once on the grain
this rapidly hydrogenates into $*$CH$_4$.
Model 4 then appears as a transitory case. It has more $*$CH$_4$ 
than model 5, about the same as the faster cases, with correspondingly
more $*$CO and $*$CO$_2$ than the faster cases and less than model 5.
It appears that for model 4 the greater time before freeze out 
for CO and CO$_2$ to form is partially offset by the lower starting density.
The same effect can be seen when comparing models 1 and 2, 
though it is not as pronounced.

$*$NH$_3$ is well modelled in models 1, 2 and 3, above the 1\% threshold in model 7 and underproduced 
in the other models. Here it seems that rapid collapse alone is the major factor. 
Abundance in model 7 is significantly lower than in models 1-3, 
entirely a consequence of the molecular hydrogen starting abundance.
Models 4-8 all predict $\geq$ 1\% $*$HNO.
Correspondingly these are the models with the least $*$NH$_3$.
The highest $*$HNO abundance is in models 7 and 8,
again a consequence of the molecular hydrogen starting abundance.
Nitrogen cannot easily form NH$_3$ in the gas or surface phase due 
to the lack of atomic hydrogen.
Instead, atomic nitrogen freezes out and combines with $*$OH 
to form $*$HNO. The lack of atomic hydrogen allows there to be  
proportionately more $*$OH available, which otherwise would 
hydrogenate to form $*$H$_2$O.
Models 4 and 5 seem to be intermediate cases, their longer collapse times
allowing some but not all of the initial gas phase atomic hydrogen 
to convert to molecular before freeze out. 

$*$HNO is not currently observed in grain mantles and
is another model prediction to be searched for. 
However its greatest abundance occurs in two collapses (models 7 and 8), 
both with high starting density.
A dense cloud with ample molecular hydrogen could have evolved 
chemically before collapse begins
reducing the $*$HNO abundance by enabling atomic nitrogen
to combine into other species in the early gas phase.
All models predict a significant mantle abundance of $*$N$_2$ and 
most models $*$O$_2$, at $\simeq$ 10\%.
In the absence of observational measurements these 
must be regarded as model predictions. 
Ehrenfreund  \& van Dishoeck (\cite{Ehrenfreund98}) suggest possible
ways of searching for $*$N$_2$ and $*$O$_2$
while Ehrenfreund \& Schutte (\cite{Ehrenfreund00})
describe the current state of observations.

Although mantle composition observations cannot be regarded as 
complete and definitive, approximate agreement for $*$H$_2$O and $*$CO
is seen in mantles produced by the Gas/Surface model.
Mantle compositions in models 3 and 6 (10,000 cm$^{-3}$ free fall 
and retarded) must be regarded as suspect as it is likely that the scenarios
themselves are unphysical, starting as they do from atomic hydrogen. 
Models 4, 5, 7 and 8 produce compositions closest to observations, 
although models 7 and 8 predict $*$CH$_3$OH abundance at the low end 
of its observed abundance range.
The effects of collapse timescale on mantle composition are shown 
in Table
\ref{Gas phase abundances - freeze out}
which lists relative abundances of gas phase C, CO, CO$_2$, H, H$_2$ and O 
with respect to total hydrogen at the beginning of freeze out.
This is arbitrarily defined, for the purposes of this discussion,
as the last timestep before at least five mantle layers form on the grain.
Models 2, 4 and particularly 5 seem to have 
an optimum combination of collapse speed and density 
to allow significant CO to form in the gas phase.
This partially depletes atomic carbon and enables freeze out to occur while
there is still some atomic hydrogen available  
with corresponding consequences for the chemistry thereafter.

\subsection{Complex molecules}
\label{Complex molecules}

\begin{table*}[ht]
\caption{Complex molecule abundances - gas phase}
\label{Complex molecule abundances - gas phase}
\centering
\begin{tabular}
{lllllll} 
\hline
& \multicolumn{6}{c}{Abundance w.r.t. H + H$_2$} \\
&
\multicolumn{2}{c}{CH$_3$OH} & 
\multicolumn{2}{c}{C$_2$H$_5$OH} &
\multicolumn{2}{c}{CH$_3$OCH$_3$} \\ 
\hline
 & Plateau Point & Plateau Point           & Plateau Point & Plateau Point           & Plateau Point & Plateau Point           \\
 &               & + 2.0 $\times 10^5$ yrs &               & + 2.0 $\times 10^5$ yrs &               & + 2.0 $\times 10^5$ yrs \\
\hline
Model 1 & 1.12 $\times 10^{-5}$   & 4.32 $\times 10^{-7}$   & 1.78 $\times 10^{-9}$   & 1.45 $\times 10^{-8}$    & 3.01 $\times 10^{-7}$   & 8.36 $\times 10^{-7}$ \\
Model 2 & 1.83 $\times 10^{-5}$   & 3.80 $\times 10^{-7}$   & 1.04 $\times 10^{-9}$   & 3.32 $\times 10^{-8}$    & 4.11 $\times 10^{-7}$   & 5.81 $\times 10^{-7}$ \\
Model 3 & 1.55 $\times 10^{-5}$   & 5.23 $\times 10^{-7}$   & 1.11 $\times 10^{-9}$   & 1.26 $\times 10^{-8}$    & 3.63 $\times 10^{-7}$   & 1.04 $\times 10^{-6}$ \\
Model 4 & 8.56 $\times 10^{-6}$   & 7.98 $\times 10^{-8}$   & 1.69 $\times 10^{-9}$   & 6.21 $\times 10^{-8}$    & 8.61 $\times 10^{-7}$   & 6.86 $\times 10^{-8} $ \\
Model 5 & 5.44 $\times 10^{-6}$   & 1.34 $\times 10^{-9}$   & 1.40 $\times 10^{-9}$   & 6.48 $\times 10^{-10}$   & 7.69 $\times 10^{-7}$   & 1.62 $\times 10^{-11}$ \\
Model 6 & 1.09 $\times 10^{-5}$   & 1.66 $\times 10^{-7}$   & 1.76 $\times 10^{-9}$   & 5.63 $\times 10^{-9}$    & 1.64 $\times 10^{-6}$   & 1.92 $\times 10^{-7}$ \\
Model 7 & 8.71 $\times 10^{-7}$   & 2.72 $\times 10^{-9}$   & 2.30 $\times 10^{-8}$   & 4.76 $\times 10^{-10}$   & 1.94 $\times 10^{-7}$   & 7.07 $\times 10^{-11}$ \\
Model 8 & 1.51 $\times 10^{-7}$   & 1.97 $\times 10^{-9}$   & 9.32 $\times 10^{-9}$   & 2.73 $\times 10^{-10}$   & 1.27 $\times 10^{-7}$   & 3.94 $\times 10^{-11}$ \\ 

\hline
\end{tabular}
\end{table*}

\begin{table*}[ht]
\caption{Molecular abundance in a sample of four low-mass protostars (Bottinelli et al. \cite{Bottinelli07})}
\label{Molecular abundance in a sample of four low-mass protostars}
\centering
\begin{tabular}{lrrrrl}
\hline\hline
Molecule         & IRAS16293-2422      & NGC1333-IRAS4A     & NGC1333-IRAS4B    & NGC1333-IRAS2A      & Ref.\\
\hline
H$_2$CO          & $1.0\times10^{-7}$   & $2.0\times10^{-8}$  & $3.0\times10^{-6}$ & $2.0\times10^{-7}$   & 1,2,3 \\
CH$_3$OH         & $1.0\times10^{-7}$   & $<1.0\times10^{-8}$ & $7.0\times10^{-7}$ & $3.0\times10^{-7}$   & 4 \\
HCOOCH$_3$-A     & $1.7\times10^{-7}$   & $3.4\times10^{-8}$  &  $1.1\times10^{-6}$ & $<6.7\times10^{-7}$ & 5,2,6 \\
HCOOH            & $6.2\times10^{-8}$   & $4.6\times10^{-9}$ & $<1.0\times10^{-6}$ & $<1.2\times10^{-7}$ & 5,2,6 \\
CH$_3$OCH$_3$    & $2.4\times10^{-7}$   & $<2.8\times10^{-8}$ & $<1.2\times10^{-6}$ & $3.0\times10^{-8}$  & 5,2,6,7 \\
CH$_3$CN         & $1.0\times10^{-8}$   & $1.6\times10^{-9}$  &  $9.5\times10^{-8}$ & $8.7\times10^{-9}$  & 5,2,6 \\
C$_2$H$_5$CN     & $1.2\times10^{-8}$   & $<1.2\times10^{-9}$ & $<7.5\times10^{-7}$ & $<1.0\times10^{-7}$ & 5,2,6 \\
\hline
\end{tabular}

\textbf{References}. 
(1) Ceccarelli et al. \cite{Ceccarelli00};
(2) Bottinelli et al. \cite{Bottinelli04};
(3) Maret et al. \cite{Maret04};
(4) Maret et al. \cite{Maret05}; \newline
(5) Cazaux et al. \cite{Cazaux};
(6) Bottinelli et al. \cite{Bottinelli07}; 
(7) J{\o}rgensen et al. \cite{Jorgensen} 
\end{table*}

\subsubsection{Complex molecules - surface phase}
\label{Complex molecules - surface phase}

The prime function of the Gas/Surface model is to investigate the
formation of complex molecules by grain surface catalysis. 
Gas phase abundances in the final stages of collapse 
are governed by surface abundances prior to mantle desorption.
The mantle abundances with respect to $*$H$_2$O of 
$*$CH$_3$OH, $*$C$_2$H$_5$OH and $*$CH$_3$OCH$_3$ at the 
greatest mantle thickness for each model are listed in Table 
\ref{Mantle composition w.r.t H$_2$O at maximum mantle thickness for each model}.
Models 1-6 all produce $*$CH$_3$OH in the 5-15\% range,
agreeing well with observations.
Models 7 and 8 produce $\simeq$ 1\%.
This is a consequence of the initial hydrogen being molecular.
$*$CH$_3$OH requires atomic hydrogen for its formation
(Section \ref{Reaction set}) and with little available 
formation is significantly reduced.

Models 1-6 all produce about the same amount of $*$C$_2$H$_5$OH
($\simeq$ 0.001\%) 
while models 7 and 8 produce significantly more ($\simeq$ 0.025\%).
For $*$CH$_3$OCH$_3$ models 1-4 and 6 produce $\simeq$ 0.0000002\%
while models 5, 7 and 8 production is higher at $\simeq$ 0.0001\%.
Formation of complex molecules
requires atomic carbon and $*$CO on the grain, the $*$CO originating 
in the gas phase (Section \ref{Reaction set}).
Maximum production takes place when the prior
gas phase has produced significant CO, while leaving sufficient
atomic carbon to form C-C-O and C-O-C structures on the surface 
after freeze out. 
At the same time if there is considerable atomic hydrogen present 
at freeze out then this will rapidly hydrogenate atomic carbon
on the grain surface leaving little available to form complex molecules.
This is demonstrated by models 7 and 8 where the initial hydrogen is entirely molecular.
Absence of much atomic hydrogen leaves CO to form preferentially in the gas phase.
It then freezes out alongside atomic carbon which cannot readily hydrogenate 
into $*$CH$_4$ and so is available to form C-C-O and C-O-C.
Models 7 and 8 show low $*$CH$_4$, enhanced $*$CO and the largest proportions
of both $*$C$_2$H$_5$OH and $*$CH$_3$OCH$_3$.
  
Models 1, 2, 3 and 6 with their faster collapse times and initial atomic hydrogen
all have significantly lower $*$CO in the mantle and hence much lower
complex molecule abundance. Models 4 and 5 have the longest collapse time of any
of the models. This leads to significantly more $*$CO, in the case of model 5
approximately the same as in the molecular hydrogen starting abundance cases.
Model 4 again appears as a transitory case. It has 15\% $*$CO, formed in the gas phase
and then frozen out. However its initial low density means that significant
atomic hydrogen is present at freeze out and this can hydrogenate the remaining
atomic carbon to $*$CH$_4$ (28\%) leaving very little for the production of complex molecules.
Model 5 avoids this situation. Here substantial $*$CO is frozen out but very little 
atomic hydrogen remains, hence the paucity of $*$CH$_4$.
Much of the remaining carbon has contributed to the high $*$CO$_2$ abundance
(also seen in models 7 and 8) however some has led to the significantly increased 
$*$CH$_3$OCH$_3$.

Each model produces a higher mantle abundance for
$*$C$_2$H$_5$OH than $*$CH$_3$OCH$_3$.
For models 1-4 and 6 the difference is greater than 3 orders of magnitude
while for models 7 and 8 it is 2 and for model 5 only 1.
The abundance difference seems to be caused by reaction channel differences.
$*$C$_2$H$_5$OH forms mainly by hydrogenation on the grain of C-C-O.
However $*$CH$_3$OCH$_3$ seems to form mainly by freeze out from the gas phase.
Its gas phase origin is a two stage process with 
C$_2$H$_6$OH$^+$ (protonated dimethyl ether) as an intermediate :

\begin{equation}
{\rm CH}_3^+ + {\rm CH}_3 {\rm OH} \rightarrow\ {\rm C}_2{\rm H}_6{\rm OH}^+ + photon 
\label{makeplus}     
\end{equation}

\begin{equation}
{\rm C}_2{\rm H}_6{\rm OH}^+  + electron \rightarrow\ {\rm CH}_3 {\rm OCH}_3 + {\rm H}   
\label{dest1}          
\end{equation}

Gas phase C$_2$H$_6$OH$^+$ can also be destroyed by :

\begin{equation}
{\rm C}_2{\rm H}_6{\rm OH}^+ + electron \rightarrow\ {\rm CH}_3 + {\rm CH}_3{\rm OH}
\label{dest2}     
\end{equation}

Essentially C$_2$H$_6$OH$^+$ is made from CH$_3$OH. 
Once formed it is then destroyed in forming CH$_3$OCH$_3$
(eqn. \ref{dest1})
which can freeze out onto the grain surface where it is stable
or it can form CH$_3$OH (eqn. \ref{dest2}) which is then available 
to form more C$_2$H$_6$OH$^+$ via eqn.  \ref{makeplus}.
As long as the temperature is low enough to ensure that there 
is a net freeze out of molecules onto the grain surface this 
process results in a build up of $*$CH$_3$OCH$_3$.
While workable this gas phase production of CH$_3$OCH$_3$
which then freezes out is less efficient than the direct 
grain surface hydrogenation which produces
$*$C$_2$H$_5$OH in significantly greater abundances.
Although production of C$_2$H$_6$OH$^+$ takes place in the gas phase 
grain surface catalysis is crucial to the process.
At the time of freeze out most CH$_3$OH is being produced
by evaporation from the grain mantle. In turn 
the dominant production of $*$CH$_3$OH is by a chain of 
direct hydrogenation reactions of $*$CO.

Comparison of reaction rates of complex molecule formation 
for the different models shows that model 5 seems to have an optimum 
combination of parameters to form $*$CH$_3$OCH$_3$,
hence its significantly higher abundance in this model.
Table \ref{Observed fractional abundances of complex molecules}
does not show any consistency in C$_2$H$_5$OH : CH$_3$OCH$_3$ ratio
and it seems from the models here that the abundance ratio 
of these two is significantly sensitive to prevailing conditions.

\subsubsection{Complex molecules - gas phase}
\label{Complex molecules - gas phase}

\begin{table}[ht]
\caption{Acceptable gas phase fractional abundance ranges for complex molecules}
\label{Acceptable gas phase fractional abundance ranges for complex molecules}
\centering
\begin{tabular}
{lll} 
\hline
Molecule      & Minimum   & Maximum \\ 
              & Abundance & Abundance \\
\hline
CH$_3$OH      & $5.0 \times 10^{-8}$   & $2.0 \times 10^{-6}$ \\
CH$_3$OCH$_3$ & $8.0 \times 10^{-9}$   & $7.4 \times 10^{-7}$ \\
\hline
\end{tabular}
\end{table}

Other than $*$CH$_3$OH no observations are currently available 
of complex molecules in grain mantles. However measurements
do exist of gas phase abundances (Table
\ref{Molecular abundance in a sample of four low-mass protostars})
and these can be compared to the results of the models
shown in Table 
\ref{Complex molecule abundances - gas phase}.
Comparisons are made at the plateau point
and at $2.0 \times 10^5$ years later,
the end point of time segment 2. 
Between these two points the model most closely resembles a hot corino.
There are uncertainties in observations and even when these can be 
minimized apparently similar sources yield significantly different 
abundance values for the same species. There are further uncertainties in the 
chemistry both in the processes of surface chemistry and the accuracy of 
gas phase rate coefficients. 
For these reasons a model-predicted gas phase abundance
is considered to be in
reasonable agreement with observations if it falls within 
half an order of magnitude of the range of observed values
at any time between the plateau point and 2.0 $\times 10^5$ years later.
The acceptable ranges for complex molecules are given in Table  
\ref{Acceptable gas phase fractional abundance ranges for complex molecules},
(derived from Table
\ref{Molecular abundance in a sample of four low-mass protostars}).

CH$_3$OH is well modelled in all scenarios, though the abundance
varies significantly.
In all of models 1-6 (all models starting with atomic hydrogen)
CH$_3$OH is overproduced at the plateau point and then 
for five of these cases falls to within
the acceptable range by $2.0 \times 10^5$ years later.
The sole exception here is model 5 (retarded collapses from 1,000 cm$^{-3}$) 
where the abundance actually drops to below the valid range $2.0 \times 10^5$ years later. 
Since the abundance has passed through the valid range during the time period
in question this is held to indicate validity. 
Models 7 \& 8 (with molecular hydrogen starting abundance)
are both within the valid range at the plateau point 
and fall below it by $2.0 \times 10^5$ years later. 
In all cases the CH$_3$OH gas phase abundance declines 
after the grain mantles are evaporated.

CH$_3$OCH$_3$ can also be considered well modelled in all cases, though its behaviour is more diverse.
Models 1, 2 \& 3 are within acceptable range at the plateau point and by 2.0 $\times$ 10$^5$ years later
abundance has actually increased, suggesting further production in the gas phase.
At this time model 2 is still within the acceptable range while 1 \& 3 overproduce.
Models 4, 5, \& 6 all overproduce slightly at the plateau point while 2.0 $\times$ 10$^5$ years later
abundances have dropped to within acceptable range for models 4 \& 6 and below acceptable range for model 5.
As with CH$_3$OH model 5 CH$_3$OCH$_3$ abundance is considered to be validly modelled 
as it has passed through the acceptable range during the relevant time period. 
In this case the preciptious drop, 4 orders of magnitude, serves to illustrate 
how volatile the chemistry becomes in the period after the grain mantles evaporate. 
Models 7 \& 8 show a similar behaviour being within acceptable range at the
plateau point and also dropping by 4 orders of magnitude 2.0 $\times$ 10$^5$ years later.

As discussed in Sections 
\ref{Mantle composition} and 
\ref{Complex molecules - surface phase}
the production of complex molecules is dependent on the gas phase abundances 
of atomic carbon and hydrogen at the point of freeze out. 
Gas phase processes are considerably better understood than surface processes,
and in particular the potential barriers used could be different 
leading to changes in the relative proportions of species.
With these caveats we conclude that complex molecules,
seen in both hot cores and hot corinos, 
can be made in the observed abundances by grain surface catalysis processes. 
Ethanol (C$_2$H$_5$OH) production has been included in the models here 
as it is frequently seen in the same environments as dimethyl ether (CH$_3$OCH$_3$)
(Table \ref{Observed fractional abundances of complex molecules}).
Currently it has not been observed in hot corinos and the
abundances modelled here must be regarded as predictions awaiting
the test of observation.
The only major distinction that can be draw between the different models
is between those starting with molecular hydrogen (models 7 \& 8)
compared to the other six. As would be expected the lack of
initial atomic hydrogen stifles the production of those molecular species 
which require it. Correspondingly this then allows higher abundances 
of other species whose \lq\lq raw material\rq\rq\ would otherwise
have been removed by combination with atomic hydrogen.
No correlation is noticable in the production of complex molecules
in either gas or surface phase with the cloud starting mass and/or collapse time.
This may be indicative of a volatile chemistry and/or of the limitations
of our current understanding.

Table \ref{Molecular abundance in a sample of four low-mass protostars}
lists observational measurements for a number of other molecules 
observed in hot corinos. These too can be compared to the models here as a validity check.
HCOOH and C$_2$H$_5$CN are not in the Gas/Surface model reaction set as used here. 

Both of these two species are obvious candidates for inclusion in future modelling.
HCOOH may be particularly significant as both gas 
(Table \ref{Molecular abundance in a sample of four low-mass protostars})
and solid phase (Table \ref{Observed interstellar ices})
data are now available for it.
Both Bottinelli et al. (\cite{Bottinelli07}) and Fuente et al. (\cite{Fuente05})
point out that HCOOH shows significantly different abundance behaviour between
corino type objects, as modelled here, and higher temperature hot cores,
unlike a majority of other observed species. The reasons for this are not known
but present an interesting issue for prospective future models.
H$_2$CO, HCOOCH$_3$ and CH$_3$CN are included and their abundances are listed in Table
\ref{Other molecule abundances - gas phase}.
The same criteria for acceptable range is used as above, namely 
a gas phase abundance is considered to be in reasonable agreement with observations 
if it falls within half an order of magnitude of the range of observed values
at any time between the plateau point and 2.0 $\times 10^5$ years later.
Relevant values are shown in 
Table \ref{Acceptable gas phase abundance ranges for other molecules}.

H$_2$CO is well modelled in all collapse scenarios.  
Table \ref{Molecular abundance in a sample of four low-mass protostars}
lists only abundance data for HCOOCH$_3$-A.
Bottinelli et al. (\cite{Bottinelli07}), from where this data is taken,
state that the abundance of the \lq\lq E\rq\rq\ form of HCOOCH$_3$
is usually very close to that of the \lq\lq A\rq\rq\ form
and that they regard the HCOOCH$_3$ total abundance to be 
twice that of HCOOCH$_3$-A.
For this reason the acceptable range of HCOOCH$_3$ is given in 
Table \ref{Acceptable gas phase abundance ranges for other molecules}
as  1.8 $\times 10^{-8}$ -  7.2 $\times 10^{-6}$,
i.e., half an order of magnitude above and below double the HCOOCH$_3$-A abundance.
Modelled HCOOCH$_3$ is underabundant at the plateau point in all cases.
By 2.0$\times 10^{5}$ years later it remains underabundant in seven of the eight cases.
Only in model 2 does it rise above the acceptable lower limit.  
CH$_3$CN is well modelled in most of the cases.
In models 1-4 \& 6 it is within the acceptable range both at the plateau point
and 2.0$\times 10^{5}$ years later. In models 5 \& 8 it is always underabundant.
In model 7 it is within acceptable range at the plateau point and has fallen
to below it by 2.0$\times 10^{5}$ years later.    
For all three of these molecules there is no descernible correlation between abundance 
produced and cloud starting mass and/or collapse time.

\section{Conclusion}
\label{Conclusion}

Overall the Gas/Surface model as used here reproduces to a reasonable
degree the observed complex molecule abundances seen in hot corinos.
In particular abundance values produced for methanol and dimethyl ether are 
comparable to those observed in both gas and surface phases. 

The surface phase molecular abundances produced by the model reflect the starting conditions, 
particularly the nature of the initial hydrogen - atomic or molecular.
This determines the degree of processing in the initial gas phase and hence the
abundance of species frozen on to the dust and available for surface chemistry.
This in turn allows an understanding of the differences in 
grain mantle composition observed in different regions.
{The model results clearly indicate that both the starting conditions 
and particularly the timescale of collapse are major contributors
to the overall chemistry.

A number of other workers in the area have reached similar conclusions.
Garrod \& Herbst (\cite{Garrod06}) specifically investigated methyl formate 
(HCOOCH$_3$). They concluded that its formation required the 
grain surface production of precursor molecules which were then released into the 
gas phase to produce methyl formate itself. This tallies with the similar behaviour
seen here although Garrod \& Herbst were able to produce abundances closer to observed values. 
They suggest a similar pattern for dimethyl ether, exactly as the Gas/Surface model shows.
Later applications of the same model produced comparable results.
Aikawa et al. (\cite{Aikawa08}) consider a collapse case where again large 
organic species (or their immediate precursors) are formed on grain mantles
before evaporation into the gas phase. Observed abundances can be both well modelled
or over and under produced depending on starting conditions and timescale.
Garrod et al. (\cite{Garrod08}) apply their model to more general star formation
cases and note specifically that the longer the time available for grain surface
chemistry to progress the more complex the eventual gas phase chemistry becomes.
In this case agreement with observations is seen for abundances of ethanol 
and dimethyl ether though not for some other complex molecules.
Hassel et al. (\cite{Hassel08}) specifically model L1529, 
a \lq\lq lukewarm\rq\rq\ corino, 
which may be the prototype of a new class of such objects.
Their analysis demonstrates that, at least in the case of L1529, 
there is some difficulty in establishing definitvely how a particular
chemical state came to evolve and a number of scenarios are possible
depending on the length of time the corino spent in each 
particular phase.

In conclusion then grain surface processing,
combined with
the variation of physical conditions modelled here,
can be regarded as a viable method for the formation of complex molecules, 
particularly ethanol and dimethyl ether, 
in the environment found
in the vicinity of a hot corino
and produce abundances comparable to those actually observed.

\begin{acknowledgements}
Astrophysics at Queens University Belfast is supported by a grant from STFC.
P. Hall wishes to thank Professor H. Fleischmann 
of Cornell University Department of Applied and Engineering Physics
for helpful conversations about optical absorbtion.
We both wish to thank an anonymous referee whose comments 
improved an earlier version of this paper.
\end{acknowledgements}

\begin{table*}[h]
\caption{Acceptable gas phase abundance ranges for other molecules}
\label{Acceptable gas phase abundance ranges for other molecules}
\centering
\begin{tabular}
{lll} 
\hline
Molecule      & Minimum   & Maximum \\ 
              & Abundance & Abundance \\
\hline
H$_2$CO       & 7.0 $\times 10^{-9}$  & 8.0 $\times 10^{-6}$ \\
HCOOCH$_3$    & 1.8 $\times 10^{-8}$  & 7.2 $\times 10^{-6}$ \\
CH$_3$CN      & 6.6 $\times 10^{-10}$ & 4.5 $\times 10^{-7}$ \\
\hline
\end{tabular}
\end{table*}

\begin{table*}[h]
\caption{Other molecule abundances - gas phase}
\label{Other molecule abundances - gas phase}
\centering
\begin{tabular}
{lllllll} 
\hline
& \multicolumn{6}{c}{Abundance w.r.t. H + H$_2$} \\
&
\multicolumn{2}{c}{H$_2$CO} & 
\multicolumn{2}{c}{HCOOCH$_3$} &
\multicolumn{2}{c}{CH$_3$CN} \\ 
\hline
 & Plateau Point & Plateau Point           & Plateau Point & Plateau Point           & Plateau Point & Plateau Point           \\
 &               & + 2.0 $\times 10^5$ yrs &               & + 2.0 $\times 10^5$ yrs &               & + 2.0 $\times 10^5$ yrs \\
\hline
Model 1 & 2.95 $\times 10^{-07}$ & 2.05 $\times 10^{-08}$ & 4.83 $\times 10^{-11}$ & 1.52 $\times 10^{-09}$ & 4.20 $\times 10^{-09}$ & 1.92 $\times 10^{-07}$ \\ 
Model 2 & 4.37 $\times 10^{-07}$ & 1.77 $\times 10^{-06}$ & 6.12 $\times 10^{-11}$ & 2.19 $\times 10^{-08}$ & 2.44 $\times 10^{-09}$ & 2.85 $\times 10^{-08}$ \\
Model 3 & 3.88 $\times 10^{-07}$ & 2.23 $\times 10^{-08}$ & 5.57 $\times 10^{-11}$ & 2.68 $\times 10^{-09}$ & 2.56 $\times 10^{-09}$ & 1.53 $\times 10^{-07}$ \\ 
Model 4 & 8.56 $\times 10^{-07}$ & 4.62 $\times 10^{-06}$ & 4.90 $\times 10^{-10}$ & 1.33 $\times 10^{-08}$ & 1.64 $\times 10^{-09}$ & 1.61 $\times 10^{-09}$ \\
Model 5 & 5.32 $\times 10^{-07}$ & 4.52 $\times 10^{-07}$ & 3.60 $\times 10^{-10}$ & 7.18 $\times 10^{-11}$ & 4.05 $\times 10^{-10}$ & 5.05 $\times 10^{-13}$ \\
Model 6 & 2.00 $\times 10^{-06}$ & 1.52 $\times 10^{-08}$ & 1.67 $\times 10^{-09}$ & 3.01 $\times 10^{-09}$ & 5.61 $\times 10^{-09}$ & 1.75 $\times 10^{-08}$ \\
Model 7 & 5.01 $\times 10^{-07}$ & 1.61 $\times 10^{-07}$ & 9.06 $\times 10^{-10}$ & 5.23 $\times 10^{-11}$ & 2.34 $\times 10^{-09}$ & 2.12 $\times 10^{-13}$ \\
Model 8 & 6.70 $\times 10^{-07}$ & 2.61 $\times 10^{-08}$ & 2.78 $\times 10^{-09}$ & 6.82 $\times 10^{-12}$ & 2.58 $\times 10^{-10}$ & 1.38 $\times 10^{-14}$ \\ 
\hline
\end{tabular}
\end{table*}


\section*{Appendix - Model Data}
\label{Appendix - Model Data}

\nopagebreak

\begin{table*}[h!]
\caption{Gas/Surface model - surface phase species set}
\label{Gas/Surface model - surface phase species set}
\centering
\begin{tabular}
{llllllllll} 
\hline
$*$C           & $*$C$_2$        & $*$C$_2$H      & $*$C$_2$H$_2$    & $*$C$_2$H$_3$ &
$*$C$_2$H$_3$O & $*$C$_2$H$_4$   & $*$C$_2$H$_5$  & $*$C$_2$H$_5$O   & $*$C$_2$H$_5$OH \\
$*$C$_2$H$_6$  & $*$C$_2$H$_6$CO & $*$C$_2$HO     & $*$C$_3$         & $*$C$_3$H &
$*$C$_3$H$_2$  & $*$C$_3$H$_3$   & $*$C$_3$H$_4$  & $*$C$_3$N        & $*$C$_3$O \\
$*$C$_4$       & $*$C$_4$H       & $*$C$_4$H$_2$  & $*$CCN           & $*$CCO &
$*$CH          & $*$CH$_2$       & $*$CH$_2$CN    & $*$CH$_2$CO      & $*$CH$_2$NH \\
$*$CH$_2$OC    & $*$CH$_2$OH     & $*$CH$_3$      & $*$CH$_3$CHO     & $*$CH$_3$CN &
$*$CH$_3$O     & $*$CH$_3$OC     & $*$CH$_3$OCH   & $*$CH$_3$OCH$_2$ & $*$CH$_3$OCH$_3$ \\
$*$CH$_3$OH    & $*$CH$_4$       & $*$CHOC        & $*$CHOH          & $*$CHOOH &
$*$CN          & $*$CO           & $*$CO$_2$      & $*$COC           & $*$COH \\
$*$H           & $*$H$_2$        & $*$H$_2$C$_3$  & $*$H$_2$CN       & $*$H$_2$CO & 
$*$H$_2$O      & $*$H$_2$O$_2$   & $*$H$_3$C$_3$N & $*$H$_3$C$_4$N    & $*$HC$_3$N \\
$*$HCN         & $*$HCO          & $*$HCOOCH$_3$  & $*$He            & $*$HNC &
$*$HNO         & $*$Mg           & $*$N           & $*$N$_2$         & $*$N$_2$O \\
$*$NH          & $*$NH$_2$       & $*$NH$_3$      & $*$NO            & $*$NO$_2$ & 
$*$O           & $*$O$_2$        & $*$O$_2$H      & $*$OCN           & $*$OH \\
\hline
\end{tabular}

\caption{Gas/Surface model - gas phase species set}
\label{Gas/Surface model - gas phase species set}
\centering
\begin{tabular}
{llllllllll} 
\hline
C                 & C$^+$             & C$^-$            & C$_2$           & C$_2^+$ &
C$_2$H            & C$_2$H$^+$        & C$_2$H$_2$       & C$_2$H$_2^+$    & C$_2$H$_3$ \\
C$_2$H$_3^+$      & C$_2$H$_3$O       & C$_2$H$_3$O$^+$  & C$_2$H$_4$      & C$_2$H$_4^+$ &
C$_2$H$_4$O$^+$   & C$_2$H$_5$        & C$_2$H$_5^+$     & C$_2$H$_5$O     & C$_2$H$_5$O$^+$ \\
C$_2$H$_5$OH      & C$_2$H$_5$OH$^+$  & C$_2$H$_6$       & C$_2$H$_6^+$    & C$_2$H$_6$CO &
C$_2$H$_6$CO$^+$  & C$_2$H$_6$O$^+$   & C$_2$H$_6$OH$^+$ & C$_2$H$_7^+$    & C$_2$H$_7$O$^+$ \\
C$_2$HO           & C$_2$HO$^+$       & C$_2$HOH$^+$     & C$_2$N$_2^+$    & C$_2$O$^+$ &
C$_3$             & C$_3^+$           & C$_3$H           & C$_3$H$^+$      & C$_3$H$_2$ \\
C$_3$H$_2^+$      & C$_3$H$_2$O$^+$   & C$_3$H$_3$       & C$_3$H$_3^+$    & C$_3$H$_4$ &
C$_3$H$_4^+$      & C$_3$H$_5^+$      & C$_3$H$_6$OH$^+$ & C$_3$N          & C$_3$N$^+$ \\
C$_3$O            & C$_3$O$^+$        & C$_4$            & C$_4^+$         & C$_4$H &
C$_4$H$^+$        & C$_4$H$_2$        & C$_4$H$_2^+$     & C$_4$H$_3^+$    & C$_4$H$_5^+$ \\
C$_4$N$^+$        & CCN               & CCN$^+$          & CCNH$^+$        & CCO &
CH                & CH$^+$            & CH$_2$           & CH$_2^+$        & CH$_2$CN \\
CH$_2$CN$^+$      & CH$_2$CO          & CH$_2$CO$^+$     & CH$_2$NH        & CH$_2$NH$_2^+$ & 
CH$_2$OC          & CH$_2$OH          & CH$_2$OH$^+$     & CH$_3$          & CH$_3^+$ \\
\hline
\end{tabular}
\end{table*}

\begin{table*}[ht]
\caption{Solid phase reactions used for complex molecule formation}
\label{Solid phase reactions used for complex molecule formation}
\centering
\begin{tabular}
{llllll} 
\hline
    & Reactant1        & Reactant2  & Product1         & Product2 & Activation       \\
    &                  &            &                  &          & Energy $E_a$ (K) \\
\hline
1   & $*$C             & $*$CO      & $*$CCO           &          & $5.0 \times 10^2$ \\   
2   & $*$C             & $*$CO      & $*$COC           &          & $5.0 \times 10^2$ \\   
3   & $*$C             & $*$H       & $*$CH            &          & 0.0 \\   
4   & $*$C             & $*$N       & $*$CN            &          & 0.0 \\   
5   & $*$C             & $*$NH      & $*$HNC           &          & 0.0 \\   
6   & $*$C             & $*$NH$_2$  & $*$H             & $*$HNC   & 0.0 \\
\hline
\end{tabular}
\end{table*}

\small

\longtab{18}{
\begin{longtable}
{llllllllrr} 
\caption{Gas phase reactions used for complex molecule destruction}\\
\label{Gas phase reactions used for complex molecule destruction}\\
\hline
 & Reactant1 & Reactant2 & Product1 & Product2 & Product3 & Product4 & $\alpha$ & $\beta$ & $\gamma$ \\
\hline
\endfirsthead
\caption{Gas phase reactions used for complex molecule destruction - continued}\\
\hline
 & Reactant1 & Reactant2 & Product1 & Product2 & Product3 & Product4 & $\alpha$ & $\beta$ & $\gamma$ \\
\hline
\endhead
\hline
\endfoot
 1   & C$^+$           & C$_2$H$_3$O   & C$_2$H$_3^+$    & CO          &        &    & $1.50 \times 10^{-09}$ & 0.00  & 0.0 \\
 2   & C$^+$           & C$_2$H$_5$O   & C$_2$H$_5^+$    & CO          &        &    & $1.50 \times 10^{-09}$ & 0.00  & 0.0 \\
 3   & C$^+$           & C$_2$HO       & C$_2$H$^+$      & CO          &        &    & $1.50 \times 10^{-09}$ & 0.00  & 0.0 \\
 4   & C$^+$           & CH$_2$OC      & C$_2$H$^+$      & HCO         &        &    & $1.50 \times 10^{-09}$ & 0.00  & 0.0 \\
 5   & C$^+$           & CH$_2$OH      & CH$_3^+$        & CO          &        &    & $1.50 \times 10^{-09}$ & 0.00  & 0.0 \\
 6   & C$^+$           & CH$_3$O       & CH$_3^+$        & CO          &        &    & $1.50 \times 10^{-09}$ & 0.00  & 0.0 \\
 7   & C$^+$           & CH$_3$OC      & C               & CH$_3^+$    & CO     &    & $1.50 \times 10^{-09}$ & 0.00  & 0.0 \\
 8   & C$^+$           & CH$_3$OCH     & CH              & CH$_3^+$    & CO     &    & $1.50 \times 10^{-09}$ & 0.00  & 0.0 \\
\end{longtable}
}

\begin{table*}
\caption{Model 1 - Free fall collapse from 100 cm$^{-3}$ - 
most abundant gas phase species}
\label{Model 1 - Free fall collapse from 100 cm$^{-3}$ - 
most abundant gas phase species}
\centering
\begin{minipage}[t]{70mm}
\begin{tabular}
{lll}
\hline
\multicolumn{3}{l}{Plateau Point}\\
\multicolumn{3}{l}{ $4.389 \times 10^6$ years }\\
\hline
  & Species & Abundance (w.r.t H + H$_2$)\\
\hline
 1 & H$_2$        & 5.000 $\times$ 10$^{-01}$ \\
 2 & He           & 1.401 $\times$ 10$^{-01}$ \\
 3 & H$_2$O       & 1.554 $\times$ 10$^{-04}$ \\
 4 & CH$_4$       & 5.260 $\times$ 10$^{-05}$ \\
 5 & CH$_3$OH     & 1.116 $\times$ 10$^{-05}$ \\
 6 & NH$_3$       & 8.369 $\times$ 10$^{-06}$ \\
 7 & N$_2$        & 6.114 $\times$ 10$^{-06}$ \\
 8 & CO           & 4.929 $\times$ 10$^{-06}$ \\
 9 & CO$_2$       & 1.352 $\times$ 10$^{-06}$ \\
10 & NH$_2$       & 3.999 $\times$ 10$^{-07}$ \\
11 & CH$_3$       & 3.254 $\times$ 10$^{-07}$ \\
12 & CH$_3$OCH$_3$& 3.006 $\times$ 10$^{-07}$ \\
13 & H$_2$CO      & 2.946 $\times$ 10$^{-07}$ \\
14 & O$_2$        & 2.416 $\times$ 10$^{-07}$ \\
15 & HCN          & 1.916 $\times$ 10$^{-07}$ \\
16 & C$_4$H       & 1.545 $\times$ 10$^{-07}$ \\
17 & HNC          & 1.193 $\times$ 10$^{-07}$ \\
18 & C$_3$H       & 8.901 $\times$ 10$^{-08}$ \\
19 & CH           & 4.796 $\times$ 10$^{-08}$ \\
20 & NH           & 2.391 $\times$ 10$^{-08}$ \\
21 & C$_3$H$_3$   & 1.873 $\times$ 10$^{-08}$ \\
22 & H            & 1.364 $\times$ 10$^{-08}$ \\
23 & CH$_2$       & 9.769 $\times$ 10$^{-09}$ \\
24 & C$_2$H$_4$   & 8.988 $\times$ 10$^{-09}$ \\
\hline
\end{tabular}
\end{minipage}
\begin{minipage}[t]{70mm}
\begin{tabular}
{lll}
\hline
\multicolumn{3}{l}{Plateau Point + 2.0 $\times 10^5$ years}\\
\multicolumn{3}{l}{ $4.589 \times 10^6$ years }\\
\hline
  & Species & Abundance (w.r.t H + H$_2$)\\
\hline
 1 & H$_2$        & 5.000 $\times$ 10$^{-01}$ \\
 2 & He           & 1.401 $\times$ 10$^{-01}$ \\
 3 & H$_2$O       & 1.447 $\times$ 10$^{-04}$ \\
 4 & CH$_4$       & 3.980 $\times$ 10$^{-05}$ \\
 5 & CO$_2$       & 1.462 $\times$ 10$^{-05}$ \\
 6 & N$_2$        & 7.593 $\times$ 10$^{-06}$ \\
 7 & HCN          & 2.847 $\times$ 10$^{-06}$ \\
 8 & HNC          & 1.607 $\times$ 10$^{-06}$ \\
 9 & C$_4$H       & 8.936 $\times$ 10$^{-07}$ \\
10 & CH$_2$CN     & 8.380 $\times$ 10$^{-07}$ \\
11 & CH$_3$OCH$_3$& 8.361 $\times$ 10$^{-07}$ \\
12 & C$_3$H$_3$   & 6.461 $\times$ 10$^{-07}$ \\
13 & HC$_3$N      & 5.801 $\times$ 10$^{-07}$ \\
14 & CH$_3$OH     & 4.315 $\times$ 10$^{-07}$ \\
15 & C$_3$H       & 3.657 $\times$ 10$^{-07}$ \\
16 & C$_2$        & 2.372 $\times$ 10$^{-07}$ \\
17 & CH$_3$CN     & 1.923 $\times$ 10$^{-07}$ \\
18 & C$_4$H$_2$   & 6.160 $\times$ 10$^{-08}$ \\
19 & CH           & 5.320 $\times$ 10$^{-08}$ \\
20 & C$_3$H$_2$   & 5.081 $\times$ 10$^{-08}$ \\
21 & C$_3$N       & 3.511 $\times$ 10$^{-08}$ \\
22 & C$_3$        & 3.456 $\times$ 10$^{-08}$ \\
23 & CH$_3$       & 2.655 $\times$ 10$^{-08}$ \\
24 & CCN          & 2.442 $\times$ 10$^{-08}$ \\
\hline
\end{tabular}
\end{minipage}
\end{table*}

\begin{table*}
\caption{Model 2 - Free fall collapse from 1,000 cm$^{-3}$ - 
most abundant gas phase species}
\label{Model 2 - Free fall collapse from 1,000 cm$^{-3}$ - 
most abundant gas phase species}
\centering
\begin{minipage}[t]{70mm}
\begin{tabular}
{lll}
\hline
\multicolumn{3}{l}{Plateau Point}\\
\multicolumn{3}{l}{ $1.388\times 10^6$ years }\\
\hline
  & Species & Abundance (w.r.t H + H$_2$)\\
\hline
 1 & H$_2$        & 5.000 $\times$ 10$^{-01}$ \\
 2 & He           & 1.401 $\times$ 10$^{-01}$ \\
 3 & H$_2$O       & 1.375 $\times$ 10$^{-04}$ \\
 4 & CH$_4$       & 3.880 $\times$ 10$^{-05}$ \\
 5 & CH$_3$OH     & 1.834 $\times$ 10$^{-05}$ \\
 6 & CO           & 8.797 $\times$ 10$^{-06}$ \\
 7 & N$_2$        & 8.063 $\times$ 10$^{-06}$ \\
 8 & NH$_3$       & 4.633 $\times$ 10$^{-06}$ \\
 9 & CO$_2$       & 3.475 $\times$ 10$^{-06}$ \\
10 & O$_2$        & 1.413 $\times$ 10$^{-06}$ \\
11 & CH$_3$       & 4.982 $\times$ 10$^{-07}$ \\
12 & H$_2$CO      & 4.366 $\times$ 10$^{-07}$ \\
13 & CH$_3$OCH$_3$& 4.110 $\times$ 10$^{-07}$ \\
14 & NH$_2$       & 3.131 $\times$ 10$^{-07}$ \\
15 & C$_4$H       & 2.115 $\times$ 10$^{-07}$ \\
16 & HCN          & 1.361 $\times$ 10$^{-07}$ \\
17 & HNC          & 8.830 $\times$ 10$^{-08}$ \\
18 & C$_3$H       & 7.771 $\times$ 10$^{-08}$ \\
19 & CH           & 3.446 $\times$ 10$^{-08}$ \\
20 & NH           & 2.147 $\times$ 10$^{-08}$ \\
21 & C$_2$H$_4$   & 1.672 $\times$ 10$^{-08}$ \\
22 & H            & 1.329 $\times$ 10$^{-08}$ \\
23 & C$_2$H$_2$   & 1.205 $\times$ 10$^{-08}$ \\
24 & OCN          & 1.130 $\times$ 10$^{-08}$ \\
\hline
\end{tabular}
\end{minipage}
\begin{minipage}[t]{70mm}
\begin{tabular}
{lll}
\hline
\multicolumn{3}{l}{Plateau Point + 2.0 $\times 10^5$ years}\\
\multicolumn{3}{l}{ $1.588\times 10^6$ years }\\
\hline
  & Species & Abundance (w.r.t H + H$_2$)\\
\hline
 1 & H$_2$        & 5.000 $\times$ 10$^{-01}$ \\
 2 & He           & 1.401 $\times$ 10$^{-01}$ \\
 3 & H$_2$O       & 1.300 $\times$ 10$^{-04}$ \\
 4 & CH$_4$       & 3.118 $\times$ 10$^{-05}$ \\
 5 & CO$_2$       & 2.121 $\times$ 10$^{-05}$ \\
 6 & N$_2$        & 8.945 $\times$ 10$^{-06}$ \\
 7 & CH$_3$       & 3.178 $\times$ 10$^{-06}$ \\
 8 & H$_2$CO      & 1.768 $\times$ 10$^{-06}$ \\
 9 & C$_4$H       & 1.450 $\times$ 10$^{-06}$ \\
10 & HCN          & 1.444 $\times$ 10$^{-06}$ \\
11 & HNC          & 1.382 $\times$ 10$^{-06}$ \\
12 & CH$_3$OCH$_3$& 5.813 $\times$ 10$^{-07}$ \\
13 & C$_2$        & 4.082 $\times$ 10$^{-07}$ \\
14 & CH$_3$OH     & 3.799 $\times$ 10$^{-07}$ \\
15 & C$_3$H       & 3.768 $\times$ 10$^{-07}$ \\
16 & CH$_2$CN     & 3.276 $\times$ 10$^{-07}$ \\
17 & C$_3$H$_3$   & 2.841 $\times$ 10$^{-07}$ \\
18 & HC$_3$N      & 2.025 $\times$ 10$^{-07}$ \\
19 & C$_3$H$_2$   & 8.748 $\times$ 10$^{-08}$ \\
20 & C$_3$        & 8.348 $\times$ 10$^{-08}$ \\
21 & CH           & 6.791 $\times$ 10$^{-08}$ \\
22 & C$_4$H$_2$   & 6.266 $\times$ 10$^{-08}$ \\
23 & CO           & 4.991 $\times$ 10$^{-08}$ \\
24 & C$_2$H$_5$OH & 3.318 $\times$ 10$^{-08}$ \\
\hline
\end{tabular}
\end{minipage}
\end{table*}

\begin{table*}
\caption{Model 3 - Free fall collapse from 10,000 cm$^{-3}$ - 
most abundant gas phase species}
\label{Model 3 - Free fall collapse from 10,000 cm$^{-3}$ - 
most abundant gas phase species}
\centering
\begin{minipage}[t]{70mm}
\begin{tabular}
{lll}
\hline
\multicolumn{3}{l}{Plateau Point}\\
\multicolumn{3}{l}{ $4.381 \times 10^5$ years }\\
\hline
  & Species & Abundance (w.r.t H + H$_2$)\\
\hline
 1 & H$_2$        & 5.000 $\times$ 10$^{-01}$ \\
 2 & He           & 1.401 $\times$ 10$^{-01}$ \\
 3 & H$_2$O       & 1.494 $\times$ 10$^{-04}$ \\
 4 & CH$_4$       & 4.801 $\times$ 10$^{-05}$ \\
 5 & CH$_3$OH     & 1.555 $\times$ 10$^{-05}$ \\
 6 & N$_2$        & 6.961 $\times$ 10$^{-06}$ \\
 7 & NH$_3$       & 6.807 $\times$ 10$^{-06}$ \\
 8 & CO           & 4.811 $\times$ 10$^{-06}$ \\
 9 & CO$_2$       & 1.700 $\times$ 10$^{-06}$ \\
10 & O$_2$        & 6.601 $\times$ 10$^{-07}$ \\
11 & CH$_3$       & 4.107 $\times$ 10$^{-07}$ \\
12 & H$_2$CO      & 3.875 $\times$ 10$^{-07}$ \\
13 & CH$_3$OCH$_3$& 3.630 $\times$ 10$^{-07}$ \\
14 & NH$_2$       & 3.204 $\times$ 10$^{-07}$ \\
15 & HCN          & 1.606 $\times$ 10$^{-07}$ \\
16 & C$_4$H       & 1.090 $\times$ 10$^{-07}$ \\
17 & HNC          & 8.841 $\times$ 10$^{-08}$ \\
18 & C$_3$H       & 6.344 $\times$ 10$^{-08}$ \\
19 & CH           & 4.268 $\times$ 10$^{-08}$ \\
20 & NH           & 2.196 $\times$ 10$^{-08}$ \\
21 & OCN          & 1.872 $\times$ 10$^{-08}$ \\
22 & C$_3$H$_3$   & 1.502 $\times$ 10$^{-08}$ \\
23 & H            & 1.335 $\times$ 10$^{-08}$ \\
24 & C$_2$H$_4$   & 1.332 $\times$ 10$^{-08}$ \\
\hline
\end{tabular}
\end{minipage}
\begin{minipage}[t]{70mm}
\begin{tabular}
{lll}
\hline
\multicolumn{3}{l}{Plateau Point + 2.0 $\times 10^5$ years}\\
\multicolumn{3}{l}{ $6.381 \times 10^5$ years }\\
\hline
  & Species & Abundance (w.r.t H + H$_2$)\\
\hline
 1 & H$_2$        & 5.000 $\times$ 10$^{-01}$ \\
 2 & He           & 1.401 $\times$ 10$^{-01}$ \\
 3 & H$_2$O       & 1.403 $\times$ 10$^{-04}$ \\
 4 & CH$_4$       & 3.849 $\times$ 10$^{-05}$ \\
 5 & CO$_2$       & 1.667 $\times$ 10$^{-05}$ \\
 6 & N$_2$        & 8.063 $\times$ 10$^{-06}$ \\
 7 & HCN          & 2.385 $\times$ 10$^{-06}$ \\
 8 & HNC          & 1.355 $\times$ 10$^{-06}$ \\
 9 & CH$_3$OCH$_3$& 1.038 $\times$ 10$^{-06}$ \\
10 & C$_4$H       & 8.916 $\times$ 10$^{-07}$ \\
11 & CH$_2$CN     & 6.885 $\times$ 10$^{-07}$ \\
12 & C$_3$H$_3$   & 5.968 $\times$ 10$^{-07}$ \\
13 & HC$_3$N      & 5.436 $\times$ 10$^{-07}$ \\
14 & CH$_3$OH     & 5.232 $\times$ 10$^{-07}$ \\
15 & C$_3$H       & 3.802 $\times$ 10$^{-07}$ \\
16 & C$_2$        & 2.579 $\times$ 10$^{-07}$ \\
17 & CH$_3$CN     & 1.528 $\times$ 10$^{-07}$ \\
18 & C$_4$H$_2$   & 6.469 $\times$ 10$^{-08}$ \\
19 & C$_3$H$_2$   & 5.613 $\times$ 10$^{-08}$ \\
20 & CH           & 5.452 $\times$ 10$^{-08}$ \\
21 & C$_3$        & 3.771 $\times$ 10$^{-08}$ \\
22 & CH$_3$       & 3.374 $\times$ 10$^{-08}$ \\
23 & C$_3$N       & 3.340 $\times$ 10$^{-08}$ \\
24 & H$_2$CO      & 2.230 $\times$ 10$^{-08}$ \\
\hline
\end{tabular}
\end{minipage}
\end{table*}

\begin{table*}
\caption{Model 4 - Retarded collapse from 100 cm$^{-3}$ - 
most abundant gas phase species}
\label{Model 4 - Retarded collapse from 100 cm$^{-3}$ - 
most abundant gas phase species}
\centering
\begin{minipage}[t]{70mm}
\begin{tabular}
{lll}
\hline
\multicolumn{3}{l}{Plateau Point}\\
\multicolumn{3}{l}{ $2.193 \times 10^7$ years }\\
\hline
  & Species & Abundance (w.r.t H + H$_2$)\\
\hline
 1 & H$_2$        & 5.000 $\times$ 10$^{-01}$ \\
 2 & He           & 1.401 $\times$ 10$^{-01}$ \\
 3 & H$_2$O       & 1.158 $\times$ 10$^{-04}$ \\
 4 & CH$_4$       & 3.190 $\times$ 10$^{-05}$ \\
 5 & CO           & 1.416 $\times$ 10$^{-05}$ \\
 6 & CO$_2$       & 1.155 $\times$ 10$^{-05}$ \\
 7 & N$_2$        & 1.033 $\times$ 10$^{-05}$ \\
 8 & CH$_3$OH     & 8.562 $\times$ 10$^{-06}$ \\
 9 & O$_2$        & 5.938 $\times$ 10$^{-06}$ \\
10 & CH$_3$OCH$_3$& 8.609 $\times$ 10$^{-07}$ \\
11 & H$_2$CO      & 8.562 $\times$ 10$^{-07}$ \\
12 & CH$_3$       & 6.489 $\times$ 10$^{-07}$ \\
13 & C$_4$H       & 5.230 $\times$ 10$^{-07}$ \\
14 & NH$_3$       & 2.774 $\times$ 10$^{-07}$ \\
15 & C$_3$H       & 1.241 $\times$ 10$^{-07}$ \\
16 & HCN          & 1.232 $\times$ 10$^{-07}$ \\
17 & HNC          & 1.008 $\times$ 10$^{-07}$ \\
18 & C$_2$H$_2$   & 8.272 $\times$ 10$^{-08}$ \\
19 & NH$_2$       & 7.967 $\times$ 10$^{-08}$ \\
20 & OCN          & 5.908 $\times$ 10$^{-08}$ \\
21 & C$_3$H$_2$   & 4.136 $\times$ 10$^{-08}$ \\
22 & C$_2$H$_4$   & 3.294 $\times$ 10$^{-08}$ \\
23 & CH           & 2.817 $\times$ 10$^{-08}$ \\
24 & C$_2$        & 2.198 $\times$ 10$^{-08}$ \\
\hline
\end{tabular}
\end{minipage}
\begin{minipage}[t]{70mm}
\begin{tabular}
{lll}
\hline
\multicolumn{3}{l}{Plateau Point + 2.0 $\times 10^5$ years}\\
\multicolumn{3}{l}{ $2.213 \times 10^7$ years }\\
\hline
  & Species & Abundance (w.r.t H + H$_2$)\\
\hline
 1 & H$_2$        & 5.000 $\times$ 10$^{-01}$ \\
 2 & He           & 1.400 $\times$ 10$^{-01}$ \\
 3 & H$_2$O       & 1.021 $\times$ 10$^{-04}$ \\
 4 & CO$_2$       & 2.822 $\times$ 10$^{-05}$ \\
 5 & CH$_4$       & 2.185 $\times$ 10$^{-05}$ \\
 6 & CO           & 1.184 $\times$ 10$^{-05}$ \\
 7 & N$_2$        & 1.034 $\times$ 10$^{-05}$ \\
 8 & H$_2$CO      & 4.623 $\times$ 10$^{-06}$ \\
 9 & C$_4$H       & 9.170 $\times$ 10$^{-07}$ \\
10 & HCN          & 2.600 $\times$ 10$^{-07}$ \\
11 & HNC          & 2.474 $\times$ 10$^{-07}$ \\
12 & C$_3$H       & 1.636 $\times$ 10$^{-07}$ \\
13 & C$_3$H$_3$   & 1.141 $\times$ 10$^{-07}$ \\
14 & CH$_3$OH     & 7.979 $\times$ 10$^{-08}$ \\
15 & C$_3$H$_2$   & 7.747 $\times$ 10$^{-08}$ \\
16 & CH$_3$OCH$_3$& 6.864 $\times$ 10$^{-08}$ \\
17 & CH$_2$CN     & 6.811 $\times$ 10$^{-08}$ \\
18 & C$_2$H$_5$OH & 6.214 $\times$ 10$^{-08}$ \\
19 & HC$_3$N      & 4.022 $\times$ 10$^{-08}$ \\
20 & C$_4$H$_2$   & 2.109 $\times$ 10$^{-08}$ \\
21 & C$_2$        & 1.836 $\times$ 10$^{-08}$ \\
22 & CH           & 1.541 $\times$ 10$^{-08}$ \\
23 & C$_3$        & 1.504 $\times$ 10$^{-08}$ \\
24 & H            & 1.501 $\times$ 10$^{-08}$ \\
\hline
\end{tabular}
\end{minipage}
\end{table*}

\begin{table*}
\caption{Model 5 - Retarded collapse from 1,000 cm$^{-3}$ - 
most abundant gas phase species}
\label{Model 5 - Retarded collapse from 1,000 cm$^{-3}$ - 
most abundant gas phase species}
\centering
\begin{minipage}[t]{70mm}
\begin{tabular}
{lll}
\hline
\multicolumn{3}{l}{Plateau Point}\\
\multicolumn{3}{l}{ $6.930 \times 10^6$ years }\\
\hline
  & Species & Abundance (w.r.t H + H$_2$)\\
\hline
 1 & H$_2$        & 5.000 $\times$ 10$^{-01}$ \\
 2 & He           & 1.400 $\times$ 10$^{-01}$ \\
 3 & H$_2$O       & 6.859 $\times$ 10$^{-05}$ \\
 4 & CO           & 4.396 $\times$ 10$^{-05}$ \\
 5 & CO$_2$       & 1.946 $\times$ 10$^{-05}$ \\
 6 & N$_2$        & 1.056 $\times$ 10$^{-05}$ \\
 7 & O$_2$        & 8.515 $\times$ 10$^{-06}$ \\
 8 & CH$_3$OH     & 5.436 $\times$ 10$^{-06}$ \\
 9 & CH$_3$OCH$_3$& 7.695 $\times$ 10$^{-07}$ \\
10 & CH$_3$       & 5.394 $\times$ 10$^{-07}$ \\
11 & H$_2$CO      & 5.316 $\times$ 10$^{-07}$ \\
12 & CH$_4$       & 4.808 $\times$ 10$^{-07}$ \\
13 & C$_2$H$_2$   & 9.598 $\times$ 10$^{-08}$ \\
14 & C$_4$H       & 6.895 $\times$ 10$^{-08}$ \\
15 & HCN          & 4.602 $\times$ 10$^{-08}$ \\
16 & NH$_3$       & 4.573 $\times$ 10$^{-08}$ \\
17 & OCN          & 3.380 $\times$ 10$^{-08}$ \\
18 & HNC          & 3.317 $\times$ 10$^{-08}$ \\
19 & C$_3$H       & 2.731 $\times$ 10$^{-08}$ \\
20 & NH$_2$       & 2.527 $\times$ 10$^{-08}$ \\
21 & C$_3$H$_2$   & 2.332 $\times$ 10$^{-08}$ \\
22 & H            & 1.423 $\times$ 10$^{-08}$ \\
23 & Mg           & 6.970 $\times$ 10$^{-09}$ \\
24 & NH           & 5.941 $\times$ 10$^{-09}$ \\
\hline
\end{tabular}
\end{minipage}
\begin{minipage}[t]{70mm}
\begin{tabular}
{lll}
\hline
\multicolumn{3}{l}{Plateau Point + 2.0 $\times 10^5$ years}\\
\multicolumn{3}{l}{ $7.130 \times 10^6$ years }\\
\hline
  & Species & Abundance (w.r.t H + H$_2$)\\
\hline
 1 & H$_2$        & 5.000 $\times$ 10$^{-01}$ \\
 2 & He           & 1.400 $\times$ 10$^{-01}$ \\
 3 & H$_2$O       & 4.986 $\times$ 10$^{-05}$ \\
 4 & CO$_2$       & 4.113 $\times$ 10$^{-05}$ \\
 5 & CO           & 2.959 $\times$ 10$^{-05}$ \\
 6 & N$_2$        & 1.064 $\times$ 10$^{-05}$ \\
 7 & O$_2$        & 6.551 $\times$ 10$^{-06}$ \\
 8 & CH$_4$       & 1.308 $\times$ 10$^{-06}$ \\
 9 & H$_2$CO      & 4.522 $\times$ 10$^{-07}$ \\
10 & C$_2$H$_4$   & 6.840 $\times$ 10$^{-08}$ \\
11 & C$_2$H$_2$   & 2.861 $\times$ 10$^{-08}$ \\
12 & O            & 2.465 $\times$ 10$^{-08}$ \\
13 & CHOOH        & 1.970 $\times$ 10$^{-08}$ \\
14 & H            & 1.893 $\times$ 10$^{-08}$ \\
15 & CH$_2$CN     & 9.355 $\times$ 10$^{-09}$ \\
16 & Mg           & 6.969 $\times$ 10$^{-09}$ \\
17 & C$_4$H       & 5.114 $\times$ 10$^{-09}$ \\
18 & OH           & 3.563 $\times$ 10$^{-09}$ \\
19 & HCN          & 3.234 $\times$ 10$^{-09}$ \\
20 & HNC          & 2.975 $\times$ 10$^{-09}$ \\
21 & N            & 2.669 $\times$ 10$^{-09}$ \\
22 & NH           & 2.406 $\times$ 10$^{-09}$ \\
23 & NH$_3$       & 1.498 $\times$ 10$^{-09}$ \\
24 & CH$_3$OH     & 1.343 $\times$ 10$^{-09}$ \\
\hline
\end{tabular}
\end{minipage}
\end{table*}

\begin{table*}
\caption{Model 6 - Retarded collapse from 10,000 cm$^{-3}$ - 
most abundant gas phase species}
\label{Model 6 - Retarded collapse from 10,000 cm$^{-3}$ - 
most abundant gas phase species}
\centering
\begin{minipage}[t]{70mm}
\begin{tabular}
{lll}
\hline
\multicolumn{3}{l}{Plateau Point}\\
\multicolumn{3}{l}{ $2.198 \times 10^6$ years }\\
\hline
  & Species & Abundance (w.r.t H + H$_2$)\\
\hline
 1 & H$_2$        & 5.000 $\times$ 10$^{-01}$ \\
 2 & He           & 1.401 $\times$ 10$^{-01}$ \\
 3 & H$_2$O       & 1.391 $\times$ 10$^{-04}$ \\
 4 & CH$_4$       & 4.169 $\times$ 10$^{-05}$ \\
 5 & CH$_3$OH     & 1.094 $\times$ 10$^{-05}$ \\
 6 & N$_2$        & 1.007 $\times$ 10$^{-05}$ \\
 7 & CO$_2$       & 7.513 $\times$ 10$^{-06}$ \\
 8 & CO           & 2.971 $\times$ 10$^{-06}$ \\
 9 & H$_2$CO      & 1.995 $\times$ 10$^{-06}$ \\
10 & O$_2$        & 1.747 $\times$ 10$^{-06}$ \\
11 & CH$_3$OCH$_3$& 1.639 $\times$ 10$^{-06}$ \\
12 & CH$_3$       & 1.302 $\times$ 10$^{-06}$ \\
13 & NH$_3$       & 5.736 $\times$ 10$^{-07}$ \\
14 & C$_4$H       & 3.498 $\times$ 10$^{-07}$ \\
15 & HCN          & 2.723 $\times$ 10$^{-07}$ \\
16 & HNC          & 1.553 $\times$ 10$^{-07}$ \\
17 & C$_3$H       & 1.509 $\times$ 10$^{-07}$ \\
18 & OCN          & 7.426 $\times$ 10$^{-08}$ \\
19 & C$_3$H$_3$   & 7.264 $\times$ 10$^{-08}$ \\
20 & NH$_2$       & 6.593 $\times$ 10$^{-08}$ \\
21 & C$_3$H$_2$   & 4.091 $\times$ 10$^{-08}$ \\
22 & CH           & 4.007 $\times$ 10$^{-08}$ \\
23 & C$_2$H$_4$   & 3.178 $\times$ 10$^{-08}$ \\
24 & C$_2$        & 2.231 $\times$ 10$^{-08}$ \\
\hline
\end{tabular}
\end{minipage}
\begin{minipage}[t]{70mm}
\begin{tabular}
{lll}
\hline
\multicolumn{3}{l}{Plateau Point + 2.0 $\times 10^5$ years}\\
\multicolumn{3}{l}{ $2.398 \times 10^6$ years }\\
\hline
  & Species & Abundance (w.r.t H + H$_2$)\\
\hline
\hline
 1 & H$_2$        & 5.000 $\times$ 10$^{-01}$ \\
 2 & He           & 1.401 $\times$ 10$^{-01}$ \\
 3 & H$_2$O       & 1.258 $\times$ 10$^{-04}$ \\
 4 & CH$_4$       & 3.347 $\times$ 10$^{-05}$ \\
 5 & CO$_2$       & 2.455 $\times$ 10$^{-05}$ \\
 6 & N$_2$        & 9.940 $\times$ 10$^{-06}$ \\
 7 & C$_4$H       & 2.133 $\times$ 10$^{-06}$ \\
 8 & HCN          & 6.115 $\times$ 10$^{-07}$ \\
 9 & C$_3$H       & 4.118 $\times$ 10$^{-07}$ \\
10 & HNC          & 3.100 $\times$ 10$^{-07}$ \\
11 & HC$_3$N      & 2.558 $\times$ 10$^{-07}$ \\
12 & C$_3$H$_3$   & 2.388 $\times$ 10$^{-07}$ \\
13 & C$_2$        & 2.148 $\times$ 10$^{-07}$ \\
14 & CH$_2$CN     & 2.026 $\times$ 10$^{-07}$ \\
15 & CH$_3$OCH$_3$& 1.920 $\times$ 10$^{-07}$ \\
16 & CH$_3$OH     & 1.656 $\times$ 10$^{-07}$ \\
17 & C$_3$H$_2$   & 9.933 $\times$ 10$^{-08}$ \\
18 & C$_4$H$_2$   & 9.856 $\times$ 10$^{-08}$ \\
19 & C$_3$        & 5.645 $\times$ 10$^{-08}$ \\
20 & CH           & 5.217 $\times$ 10$^{-08}$ \\
21 & C$_3$N       & 2.139 $\times$ 10$^{-08}$ \\
22 & CH$_3$CN     & 1.751 $\times$ 10$^{-08}$ \\
23 & CH$_3$       & 1.596 $\times$ 10$^{-08}$ \\
24 & H$_2$CO      & 1.518 $\times$ 10$^{-08}$ \\
\hline
\end{tabular}
\end{minipage}
\end{table*}

\begin{table*}
\caption{Model 7 - Free fall collapse from 10,000 cm$^{-3}$ -
H$_2$ starting abundance - most abundant gas phase species}
\label{Model 7 - Free fall collapse from 10,000 cm$^{-3}$ -
H$_2$ starting abundance - most abundant gas phase species}
\centering
\begin{minipage}[t]{70mm}
\begin{tabular}
{lll}
\hline
\multicolumn{3}{l}{Plateau Point}\\
\multicolumn{3}{l}{ $4.381 \times 10^5$ years }\\
\hline
  & Species & Abundance (w.r.t H + H$_2$)\\
\hline
 1 & H$_2$        & 5.000 $\times$ 10$^{-01}$ \\
 2 & He           & 1.400 $\times$ 10$^{-01}$ \\
 3 & H$_2$O       & 8.079 $\times$ 10$^{-05}$ \\
 4 & CO           & 3.734 $\times$ 10$^{-05}$ \\
 5 & CO$_2$       & 2.562 $\times$ 10$^{-05}$ \\
 6 & N$_2$        & 1.054 $\times$ 10$^{-05}$ \\
 7 & CH$_4$       & 2.491 $\times$ 10$^{-06}$ \\
 8 & O$_2$        & 2.137 $\times$ 10$^{-06}$ \\
 9 & CH$_3$OH     & 8.713 $\times$ 10$^{-07}$ \\
10 & C$_4$H       & 7.661 $\times$ 10$^{-07}$ \\
11 & C$_3$H       & 5.091 $\times$ 10$^{-07}$ \\
12 & H$_2$CO      & 5.005 $\times$ 10$^{-07}$ \\
13 & CH$_3$OCH$_3$& 1.936 $\times$ 10$^{-07}$ \\
14 & C$_2$H$_2$   & 1.038 $\times$ 10$^{-07}$ \\
15 & CCN          & 5.875 $\times$ 10$^{-08}$ \\
16 & CH$_3$       & 5.190 $\times$ 10$^{-08}$ \\
17 & HCN          & 4.327 $\times$ 10$^{-08}$ \\
18 & C$_3$        & 3.276 $\times$ 10$^{-08}$ \\
19 & NH$_3$       & 2.411 $\times$ 10$^{-08}$ \\
20 & C$_2$        & 2.391 $\times$ 10$^{-08}$ \\
21 & C$_3$H$_2$   & 2.371 $\times$ 10$^{-08}$ \\
22 & HNC          & 2.314 $\times$ 10$^{-08}$ \\
23 & C$_2$H$_5$OH & 2.301 $\times$ 10$^{-08}$ \\
24 & HC$_3$N      & 1.915 $\times$ 10$^{-08}$ \\
\hline
\end{tabular}
\end{minipage}
\begin{minipage}[t]{70mm}
\begin{tabular}
{lll}
\hline
\multicolumn{3}{l}{Plateau Point + 2.0 $\times 10^5$ years}\\
\multicolumn{3}{l}{ $6.381 \times 10^5$ years }\\
\hline
  & Species & Abundance (w.r.t H + H$_2$)\\
\hline
 1 & H$_2$        & 5.000 $\times$ 10$^{-01}$ \\
 2 & He           & 1.400 $\times$ 10$^{-01}$ \\
 3 & H$_2$O       & 5.528 $\times$ 10$^{-05}$ \\
 4 & CO$_2$       & 4.766 $\times$ 10$^{-05}$ \\
 5 & CO           & 2.288 $\times$ 10$^{-05}$ \\
 6 & N$_2$        & 1.064 $\times$ 10$^{-05}$ \\
 7 & CH$_4$       & 1.801 $\times$ 10$^{-06}$ \\
 8 & O$_2$        & 8.088 $\times$ 10$^{-07}$ \\
 9 & H$_2$CO      & 1.609 $\times$ 10$^{-07}$ \\
10 & O            & 2.585 $\times$ 10$^{-08}$ \\
11 & C$_2$H$_2$   & 2.445 $\times$ 10$^{-08}$ \\
12 & C$_4$H       & 2.276 $\times$ 10$^{-08}$ \\
13 & CHOOH        & 2.094 $\times$ 10$^{-08}$ \\
14 & CH$_2$CN     & 2.088 $\times$ 10$^{-08}$ \\
15 & H            & 1.890 $\times$ 10$^{-08}$ \\
16 & C$_2$H$_4$   & 7.726 $\times$ 10$^{-09}$ \\
17 & Mg           & 6.969 $\times$ 10$^{-09}$ \\
18 & C$_3$H       & 4.168 $\times$ 10$^{-09}$ \\
19 & OH           & 3.816 $\times$ 10$^{-09}$ \\
20 & CH$_3$OH     & 2.722 $\times$ 10$^{-09}$ \\
21 & N            & 2.559 $\times$ 10$^{-09}$ \\
22 & NH           & 2.388 $\times$ 10$^{-09}$ \\
23 & NH$_3$       & 1.793 $\times$ 10$^{-09}$ \\
24 & HCN          & 1.226 $\times$ 10$^{-09}$ \\
\hline
\end{tabular}
\end{minipage}
\end{table*}

\begin{table*}
\caption{Model 8 - Retarded collapse from 10,000 cm$^{-3}$ -
H$_2$ starting abundance - most abundant gas phase species}
\label{Model 8 - Retarded collapse from 10,000 cm$^{-3}$ -
H$_2$ starting abundance - most abundant gas phase species}
\centering
\begin{minipage}[t]{70mm}
%
\vspace{-20mm}
\caption {Model 1 - Free fall collapse from 100 cm$^{-3}$ - gas phase data}
\label{Model 1 - Free fall collapse from 100 cm$^{-3}$ - gas phase data}
\end{figure}

\clearpage

\begin{figure}[ht]
\setlength{\unitlength}{1mm}
%
\vspace{-20mm}
\caption {Model 2 - Free fall collapse from 1,000 cm$^{-3}$ - gas phase data}
\label{Model 2 - Free fall collapse from 1,000 cm$^{-3}$ - gas phase data}
\end{figure}

\clearpage

\begin{figure}[ht]
\setlength{\unitlength}{1mm}
%
\vspace{-20mm}
\caption {Model 3 - Free fall collapse from 10,000 cm$^{-3}$ - gas phase data}
\label{Model 3 - Free fall collapse from 10,000 cm$^{-3}$ - gas phase data}
\end{figure}

\clearpage

\begin{figure}[ht]
\setlength{\unitlength}{1mm}
%
\vspace{-20mm}
\caption {Model 4 - Retarded collapse from 100 cm$^{-3}$ - gas phase data}
\label{Model 4 - Retarded collapse from 100 cm$^{-3}$ - gas phase data}
\end{figure}

\clearpage

\begin{figure}[ht]
\setlength{\unitlength}{1mm}
%
\vspace{-20mm}
\caption {Model 5 - Retarded collapse from 1,000 cm$^{-3}$ - gas phase data}
\label{Model 5 - Retarded collapse from 1,000 cm$^{-3}$ - gas phase data}
\end{figure}

\clearpage

\begin{figure}[ht]
\setlength{\unitlength}{1mm}
%
\vspace{-20mm}
\caption {Model 6 - Retarded collapse from 10,000 cm$^{-3}$ - gas phase data}
\label{Model 6 - Retarded collapse from 10,000 cm$^{-3}$ - gas phase data}
\end{figure}

\clearpage

\begin{figure}[ht]
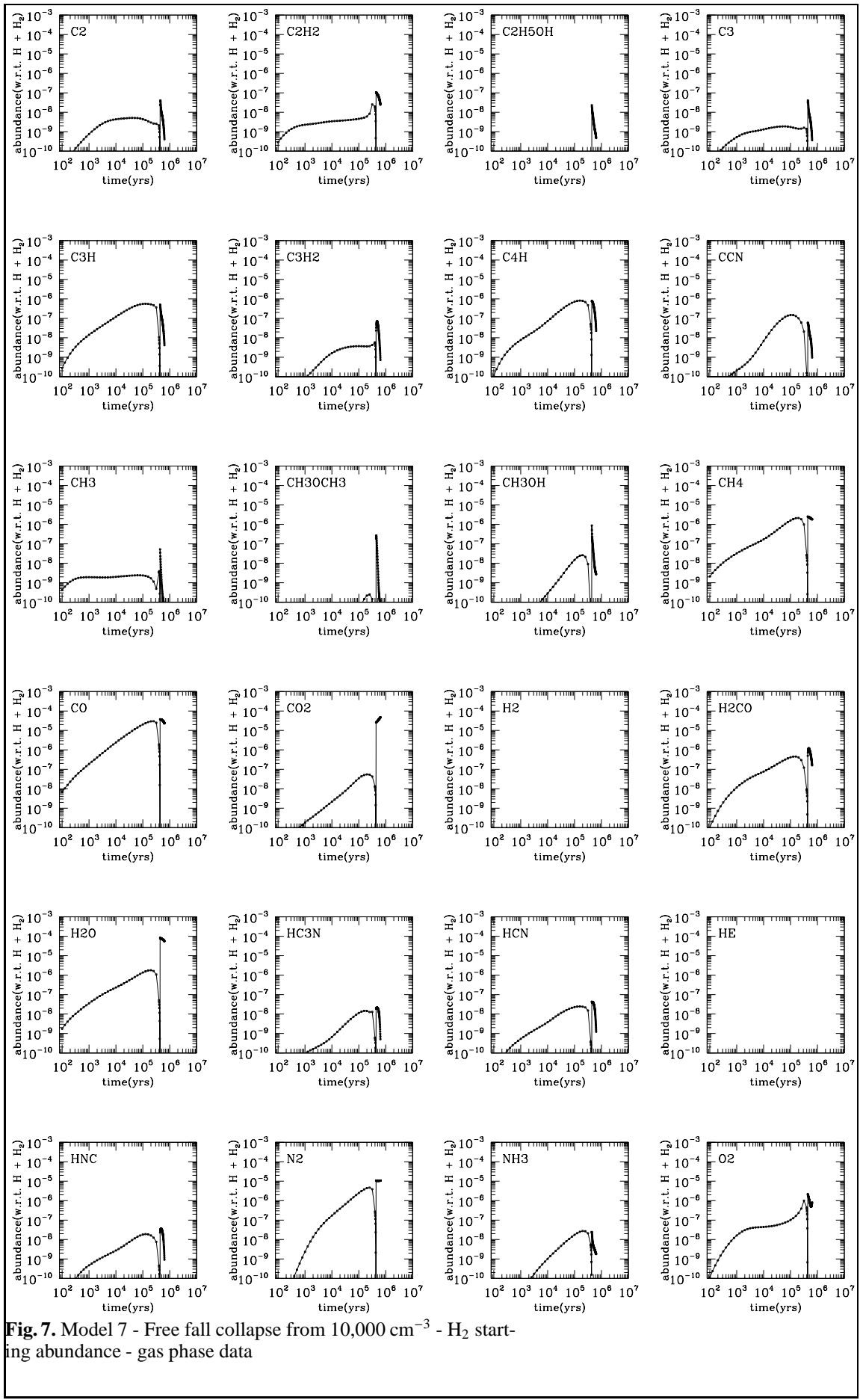

\setlength{\unitlength}{1mm}
%
\vspace{-20mm}
\caption {Model 7 - Free fall collapse from 10,000 cm$^{-3}$ - H$_2$ starting abundance - gas phase data}
\label{Model 7 - Free fall collapse from 10,000 cm$^{-3}$ - H$_2$ starting abundance - gas phase data}
\end{figure}

\clearpage

\begin{figure}[ht]
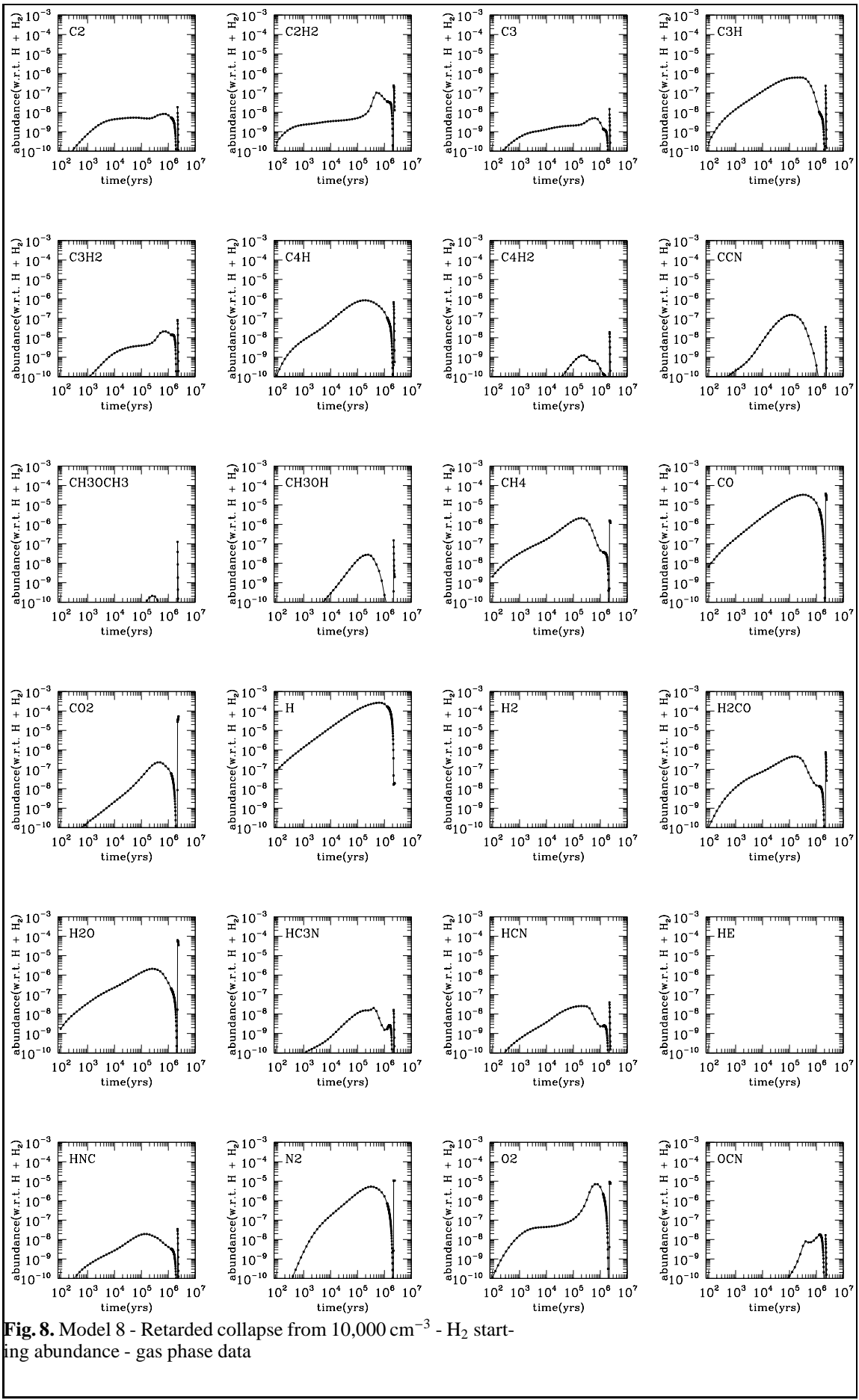

\setlength{\unitlength}{1mm}
%
\vspace{-20mm}
\caption {Model 8 - Retarded collapse from 10,000 cm$^{-3}$ - H$_2$ starting abundance - gas phase data}
\label{Model 8 - Retarded collapse from 10,000 cm$^{-3}$ - H$_2$ starting abundance - gas phase data}
\end{figure}

\clearpage



\begin{thebibliography}{}

\bibitem[2008]{Aikawa08} Aikawa Y. et al. 2008,
\apj\ 674, 993

\bibitem[1977]{Allen} Allen M. \& Robinson G.W. 1977,
\apj\ 212, 396

\bibitem[1970]{Ball} Ball J.A. 1970,
\apjl\ 162, L203

\bibitem[1996]{Boogert96} Boogert A.C.A. et al. 1996,
\aap\ 315, L377

\bibitem[2008]{Boogert08} Boogert A.C.A. et al. 2008,
\apj\ 678, 985

\bibitem[2004]{Bottinelli04} Bottinelli S., Ceccarelli C., Lefloch B., et al. 2004
\apj\ 615, 354

\bibitem[2007]{Bottinelli07} Bottinelli S., Ceccarelli C., Williams J. P. \& Lefloch B. 2007
\aap\ 463, 601

\bibitem[1991]{Breukers} Breukers R.J.L.H. 1991,
Ph.D. Thesis University of Leiden

\bibitem[1988]{Brown} Brown P.D. 1988,
Ph.D. Thesis University of Manchester

\bibitem[1992]{Carral} Carral P. \& Welch W.J. 1992,
\apj\ 385, 244

\bibitem[1993]{Caselli} Caselli P., Hasegawa T.I. \& Herbst E. 1993,
\apj\ 408, 548

\bibitem[2003]{Cazaux} Cazaux S., Tielens A.G.G.M., Ceccarelli C., et al. 2003, 
\apjl\ 593, L51

\bibitem[2000]{Ceccarelli00} Ceccarelli C., Loinard, L., Castets, A., Tielens, A.G.G.M., \& Caux, E. 2000, 
\aap\ 357, L9

\bibitem[2004]{Ceccarelli04} Ceccarelli C. 2004.
In ASP Conf. Ser. 323: Star Formation in the Interstellar Medium, 195,
ed. Johnstone D., Adams F.C., Lin F.N.C, Neufeld D.A., \& Ostriker E.C. 
Astronomical Society of the Pacific

\bibitem[1992]{Charnley92} Charnley S.B., Tielens A.G.G.M. \& Millar T.J. 1992,
\apjl\ 399, L71

\bibitem[1995]{Charnley95} Charnley S.B. et al. 1995,
\apj\ 448, 232

\bibitem[1995]{De Pree} De Pree C.G., Rodriguez L.F., Dickel H.R. 
\& Goss W. M. 1995,
\apj\ 447, 220

\bibitem[1978]{Draine} Draine B.T. 1978,
\apjs\ 36, 595

\bibitem[1984]{Duley} Duley W.W. \& Williams D.A. 1984,
Interstellar Chemistry, Academic Press.

\bibitem[1980]{Dyson} Dyson J.E. \& Williams D.A. 1980,
Physics of the Interstellar Medium, 
Manchester University Press

\bibitem[1998]{Ehrenfreund98} Ehrenfreund P. \& van Dishoeck E.F. 1998,
Advances in Space Research, 21, 15-20. 

\bibitem[2000]{Ehrenfreund00} Ehrenfreund P. \& Schutte W.A. 2000,
Advances in Space Research, 25, 2177-2188. 

\bibitem[2005]{Fuente05} Fuente A., Neri R. \& Caselli P. 2005,
\aap\ 444, 481

\bibitem[2006]{Garrod06} Garrod R.T. \&  Herbst E. 2006,
\aap\ 457, 927

\bibitem[2007]{Garrod07} Garrod R.T., Wakelam V. \&  Herbst E. 2007,
\aap\ 467, 1103

\bibitem[2008]{Garrod08} Garrod R.T., Widicus Weaver S.L. \&  Herbst E. 2008,
\apj\ 682, 283

\bibitem[2009]{Garrod09} Garrod R.T., Vasyunin A.I., Semenov D.A., Wiebe D.S., Henning Th. 2009,
\apjl\ 700, L43

\bibitem[2000]{Gibb} Gibb E. et al. 2000,
\apj\ 536, 347

\bibitem[1986]{Grim} Grim R.J.A. \& d'Hendecourt L.B. 1986,
\aap\ 167, 161

\bibitem[1997]{Hall} Hall P. 1997.
In Dust and Molecules in Evolved Stars, 203, 
ed. Cherchneff I. \& Millar T.J. 
Kluwer Academic Publishers

\bibitem[1992]{Hasegawa92} Hasegawa T.I., Herbst E. \& Leung C.M. 1992,
\apjs\ 82, 167

\bibitem[1993]{Hasegawa93} Hasegawa T.I. \& Herbst E. 1993,
\mnras\ 261, 83

\bibitem[2008]{Hassel08} Hassel G.E., Herbst E. \& Garrod R.T. 2008,
\apj\ 681, 1385

\bibitem[1989]{Herbst} Herbst E. \& Leung C.M. 1989,
\apjs\ 69, 271

\bibitem[1974]{Hindmarsh74} Hindmarsh A.C. 1974,
Lawrence Livermore Laboratory
Rept. ucrl-30001 Rev.3

\bibitem[1983]{Hindmarsh83} Hindmarsh A.C. 1983,
In Scientific Computing, 55,
ed. Stepleman R.S. et al. 
North-Holland Amsterdam

\bibitem[1971]{Hollenbach71} Hollenbach D.J. \& Salpeter E.E. 1971.
\apj\ 163, 155.

\bibitem[2001]{Ikeda01} Ikeda M. et al. 2001,
\apj\ 560, 792

\bibitem[2002]{Ikeda02} Ikeda M. et al. 2002,
\apj\ 571, 560

\bibitem[2005]{Jorgensen} J{\o}rgensen J.K., Bourke T.L., Myers P.C., et al. 2005
\apj\ 632, 973

\bibitem[1998]{Lacy} Lacy J.H., Faraji H., Sandford S.A. \& Allamandola L.J. 1998,
\apj\ 501, L105

\bibitem[1984]{Leung} Leung C.M., Herbst E. \& Huebner W.F. 1984,
\apjs\ 56, 231

\bibitem[2004]{Maret04} Maret S., Ceccarelli C., Caux E., et al. 2004, 
\aap\ 416, 577

\bibitem[2005]{Maret05} Maret S., Ceccarelli C., Tielens A.G.G.M., et al. 2005, 
\aap\ 442, 527

\bibitem[1990]{Millar90} Millar T.J. 1990,
In Molecular Astrophysics, 115,
ed. Hartquist T.W. University of Cambridge Press.

\bibitem[1991a]{Millar91a} Millar T.J. et al. 1991,
\aaps\ 87, 585

\bibitem[1991b]{Millar91b} Millar T.J., Herbst E. \& Charnley S.B. 1991,
\apj\ 369, 147

\bibitem[1997]{Millar97} Millar T.J., Farquhar P.R.A. \& Willacy K. 1997,
\aaps\ 121, 139

\bibitem[1996]{Mumma} Mumma M.J. et al. 1996,
Sci. 272, 1310

\bibitem[2008]{Oberg08} $\stackrel{..}{\rm O}$berg K.I. et al. 2008,
\apj\ 678, 1032

\bibitem[2008]{Pontoppidan} Pontoppidan K.M. et al. 2008,
\apj\ 678, 1005

\bibitem[2003]{Remijan} Remijan A. et. al. 2003,
\apj\ 590, 314

\bibitem[1988]{Schutte} Schutte W.A. \& Greenberg J.M. 1988.
In Dust in the Universe, 403.\\
eds. Bailey M.E. \& Williams D.A.
Cambridge University Press.

\bibitem[1995]{Shalabiea95a} Shalabiea O.M. 1995.
Ph.D. Thesis University of Leiden.

\bibitem[1995]{Shalabiea95b} Shalabiea O.M. \& Greenberg J.M. 1995,
\aap\ 303, 233

\bibitem[1993]{Shu} Shu F. et al. 1993,
In Protostars and Planets III, 3, 
eds. Levy E.H. \& Lunine J.I.  
University of Arizona Press

\bibitem[1974]{Snyder} Snyder L. E. et al. 1974,
\apjl\ 191, L79

\bibitem[1978]{Spitzer78} Spitzer L.J. 1978,
Physical Processes in the Interstellar Medium,
Wiley International

\bibitem[1973]{Spitzer73} Spitzer L.J. \& Cochran W.D. 1973,
\apjl\ 186, L23.

\bibitem[1996]{Tapia} Tapia M., Persi P. \& Roth M. 1996,
\aap\ 316, 102

\bibitem[1985]{Tarafdar} Tarafdar S.P. et al. 1985,
\apj\ 289, 220

\bibitem[1982]{Tielens82} Tielens A.G.G.M. \& Hagen W. 1982,
\aap\ 114, 245

\bibitem[1987]{Tielens87} Tielens A.G.G.M. \& Allamandola L.J. 1987,
In Interstellar Processes, 397,
eds. Hollenbach D.J. \& Thronson H.A.
Dordrecht Kluwer

\bibitem[1997]{Tielens97} Tielens A.G.G.M. \& Charnley S.B. 1997,
Origins of Life and Evoloution of the Biosphere, 27, 34.

\bibitem[1986]{Dishoeck} van Dishoeck E.F. \& Black J.C. 1986, 
\apjs\ 62, 109

\bibitem[1980]{Volk} V$\stackrel{..}{\rm o}$lk et al. 1980,
\aap\ 85, 316

\bibitem[1992]{Wagenblast} Wagenblast R. 1992,
\mnras\ 259, 155

\bibitem[1992]{Whittet92} Whittet D.C.B. 1992,
Dust in the Galatic Environment,
Institute of Physics Publishing.

\bibitem[1993]{Willacy93} Willacy K. 1993,
Ph.D. Thesis University of Manchester

\bibitem[1998]{Willacy98} Willacy K. \& Millar T.J. 1998,
\mnras\ 298, 562

\bibitem[1975]{Zuckerman} Zuckerman B. et al. 1975,
\apjl\ 196, L99

\end{thebibliography}
\end{document}